\documentclass[preprint,12pt,numbers,sort&compress]{elsarticle}

\usepackage{amssymb}
\usepackage{amsmath}
\usepackage[hidelinks]{hyperref}
\usepackage{dsfont,mathtools}

\usepackage{xcolor} 
\definecolor{revblue}{RGB}{0,0,150}
\newcommand{\rev}[1]{\textcolor{revblue}{#1}}
\renewcommand{\rev}[1]{#1}

\journal{Chemical Physics}

\allowdisplaybreaks
\begin{document}

\begin{frontmatter}



\title{Spin dynamics and ortho--para conversion in H\textsubscript{2}O \rev{during the phase transition from gas to solid} in external magnetic fields}


 \author[kannis]{Chrysovalantis~S.~Kannis\corref{cor1}} 
 \cortext[cor1]{Corresponding author}
 \ead{chrysovalantis.kannis@uni-duesseldorf.de}
 \author[engels1,engels2]{Ralf~Engels}
 \author[engels1,engels2,faatz3]{Nicolas~Faatz} 
 \author[engels1,engels2,puetz3]{Simon~J.~P{\"u}tz} 
  \author[kannis,buescher2]{Markus~B{\"u}scher}
  
\affiliation[kannis]{organization={Institut f{\"u}r Laser- und Plasmaphysik, Heinrich-Heine-Universit{\"a}t D{\"u}sseldorf},
	addressline={Universit{\"a}tsstra{\ss}e~1},
	city={D{\"u}sseldorf},
	postcode={40225},
	state={NRW},
	country={Germany}}

\affiliation[engels1]{organization={GSI Helmholtzzentrum f{\"u}r Schwerionenforschung},
	addressline={Planckstra{\ss}e~1}, 
	city={Darmstadt},
	postcode={64291}, 
	state={Hessen},
	country={Germany}}
	
\affiliation[engels2]{organization={Insitut f{\"u}r Kernphysik, Forschungszentrum J{\"u}lich},
	addressline={Wilhelm-Johnen-Stra{\ss}e~1}, 
	city={J{\"u}lich},
	postcode={52428}, 
	state={NRW},
	country={Germany}}

\affiliation[faatz3]{organization={III.~Physikalisches Institut B, RWTH Aachen University},
	addressline={Otto-Blumenthal-Stra{\ss}e~19}, 
	city={Aachen},
	postcode={52074}, 
	state={NRW},
	country={Germany}}

\affiliation[puetz3]{organization={Institut f{\"u}r Kernphysik, Universit{\"a}t zu K{\"o}ln},
	addressline={Z{\"u}lpicher Stra{\ss}e~77}, 
	city={K{\"o}ln},
	postcode={50937}, 
	state={NRW},
	country={Germany}}

\affiliation[buescher2]{organization={Peter-Gr{\"u}nberg Institut, Forschungszentrum J{\"u}lich},
	addressline={Wilhelm-Johnen-Stra{\ss}e~1}, 
	city={J{\"u}lich},
	postcode={52428}, 
	state={NRW},
	country={Germany}}
	

\begin{abstract}
The spin dynamics of water ice in the presence of external magnetic fields are investigated. The employed model \rev{is based on} the approach introduced by Buntkowsky et al.~[Z.~Phys.~Chem.~222, 1049 (2008)], which considers two nearest-neighbor water molecules and yields a four-spin system, as the abundant oxygen isotope has zero nuclear spin.  The model is extended to include coupling to external magnetic fields, allowing us to analyze the interplay between magnetic dipole--dipole interactions and magnetic field coupling. Two types of configurations are examined: (i) static, homogeneous fields, corresponding to a time-independent interaction, and (ii) spatially varying sinusoidal fields in relative motion with the molecules, leading to a time-dependent interaction. All computations are performed within the density operator formalism. The ortho/para populations and the total spin projections are evaluated during the first tens of milliseconds following the gas-to-solid phase transition. For static homogeneous fields, we show that increasing field strength suppresses dipolar-induced depolarization. Assuming that all molecules are initially in the para state, we show that static homogeneous fields can drive the ortho population up to approximately $50\%$, whereas suitably chosen sinusoidal-field configurations can increase it beyond $90\%$. These results are relevant for schemes aiming to preserve or manipulate nuclear-spin polarization during deposition.
\end{abstract}

\begin{graphicalabstract}
\includegraphics[width=1.0\columnwidth]{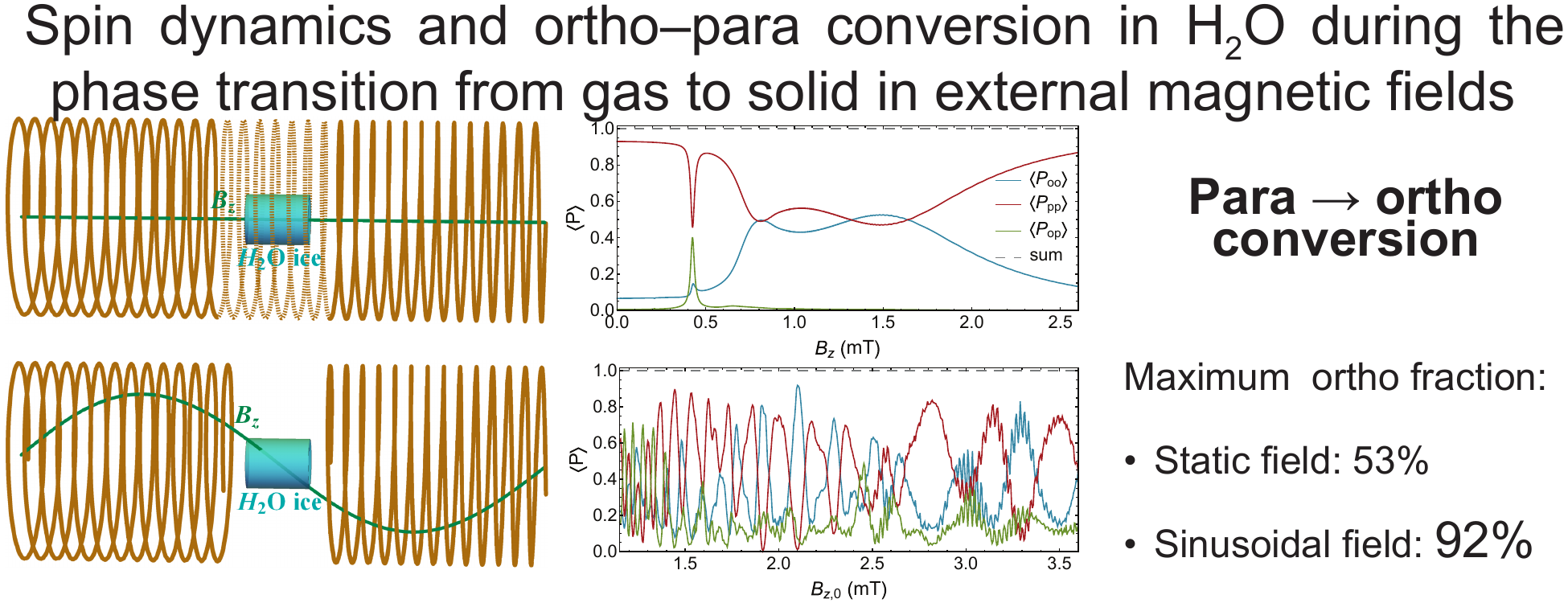}
\end{graphicalabstract}

\begin{highlights}
\item Spin dynamics of water ice in external magnetic fields.
\item Magnetic-field influence on the ortho--para conversion in water ice.
\item Preservation of nuclear-spin polarization against dipolar depolarization.
\item Implications for cryogenic deposition of nuclear-spin polarized H\textsubscript{2}O beams.
\end{highlights}

\begin{keyword}
water ice \sep nuclear-spin dynamics \sep dipolar interactions \sep ortho--para conversion \sep cryogenic deposition \sep polarization preservation


\end{keyword}

\end{frontmatter}



\section{Introduction}
\label{intro}

Water (H\textsubscript{2}O) is one of the most abundant and scientifically important molecules. It plays a central role in disciplines such as biology, physics, chemistry, and medicine. In astrophysical environments---including planets, comets, and interstellar clouds---H\textsubscript{2}O is one of the most commonly observed molecules, especially in its solid phase~\cite{leger1979,weaver1987,mumma1987,cernicharo1996}. A particularly interesting quantum property of H\textsubscript{2}O arises from the exchange symmetry, or better exchange degeneracy, associated with its two identical protons. Since protons are spin-$\frac{1}{2}$ particles, they obey Fermi–Dirac statistics. As a result, the total molecular wave function---including all relevant parts such as electronic, vibrational, rotational, spin components, etc.---must be antisymmetric under proton exchange. The spin function for a two-proton system is symmetric in the triplet state (total spin $1$) and antisymmetric in the singlet state (total spin $0$). The species of the water with the first nuclear spin symmetry is known as ortho-water, while the second is called para-water. In the ground electronic and vibrational state of a gas-phase molecule, the symmetry of the total wave function dictates that ortho-water is associated with antisymmetric rotational levels $J$, while para-water corresponds to symmetric $J$ levels. This correspondence leads to an additional $3{:}1$ statistical weight in favor of the rotational levels corresponding to ortho states, due to the total nuclear spin multiplicity.

Transitions between rotational levels of ortho and para species have a negligibly small probability and are therefore considered ``forbidden'' for all practical purposes---see, for example, Refs.~\cite{raich1964,dodelson1986} for analogous behavior in H\textsubscript{2}. For this reason, the ortho/para ratio of H\textsubscript{2}O is often used as a diagnostic tool, for instance to infer the surface temperature of cometary ice~\cite{mumma1987}. However, the effects of lattice bonding in ice and of the sublimation process on the ortho/para ratio remain less thoroughly explored. Buntkowsky et al.~\cite{buntkowsky2008} proposed a minimal theoretical model accounting for ortho--para spin conversion in water ice. According to their results, spin conversion proceeds very fast within the solid lattice. In contrast, after sublimation into the gas phase, the water molecules are effectively isolated; the rotational degrees of freedom dominate the spin interactions in the internal Hamiltonian, and the energy splitting between ortho and para states becomes large, rendering further spin conversion in the gas phase practically impossible~\cite{miani2004}. \rev{In addition, ortho--para conversion has been studied experimentally for water molecules encapsulated in C\textsubscript{60} fullerenes, where each cage contains a single H\textsubscript{2}O molecule and suppresses proton exchange between neighboring molecules while largely preserving rotational freedom~\cite{meier2018}. Measurements on  H\textsubscript{2}O@C\textsubscript{60} demonstrated para-to-ortho conversion on timescales of several tens of seconds and provided a unique model system for studying nuclear-spin isomer conversion in isolated water molecules.}

Besides the aforementioned applications in astrophysics, which focus on the spin dynamics during the naturally occurring phase transition of water from solid to gas, several laboratory-based applications involve the inverse phase transition, namely from gas to solid. In particular, a wide range of methods have been developed to polarize the nuclear spins of atomic or molecular systems in beam form, i.e., in the gas phase. However, subsequent handling of these systems may require their freezing onto a cold surface for storage and/or transportation~\cite{kannis2021, kannis20252}, enabling their use in nuclear magnetic resonance spectroscopy~\cite{edwards1986} or as fuel in polarized fusion~\cite{ciullo2016, baylor2023}. 

In view of these applications, we employ the model of Buntkowsky et al.~\cite{buntkowsky2008} to investigate the spin dynamics during the freezing of water molecules into a crystal ice lattice. We extend this model by incorporating interactions with external magnetic fields. Using the density operator formalism, we describe the spin dynamics in the H\textsubscript{2}O ice lattice and analyze how spin-related observables are influenced by the combined action of internal spin couplings and applied external fields. \rev{This work demonstrates that externally applied magnetic fields can be used as a control parameter for both depolarization and ortho--para conversion in water ice.} We identify field regimes that suppress dipolar depolarization during the phase transition of an initially nuclear-spin-polarized beam \rev{and show that suitably chosen time-dependent field configurations can enhance ortho--para conversion from an initially para beam. These findings provide guidance for future experiments aimed at preserving or manipulating nuclear-spin polarization during cryogenic deposition processes.}

This work is structured as follows. Section~\ref{theorya} introduces the model in the absence of external perturbations, followed by a discussion of the formalism and relevant quantities. In Section~\ref{theoryb}, a static homogeneous magnetic field is included, while Section~\ref{theoryc} extends the analysis to a time-dependent case by considering a spatially varying sinusoidal field in relative motion with the system. This configuration is chosen for its demonstrated effectiveness in spin manipulation and nuclear spin polarization enhancement in atomic and molecular beams~\cite{kannis2025}. The main results are discussed and conclusions are drawn in Section~\ref{conclusion}.

\section{Minimal model in the absence of external fields}
\label{theorya}

In the gas phase, H\textsubscript{2}O behaves as a quantum-mechanical free rotor, specifically, an asymmetric-top rotor. Its rotational levels are commonly denoted by $J_{K_{a} K_{c}}$, which represent $2J+1$ distinct rotational sublevels for each value of $J$ (not to be confused with the multiplicity of rotational angular momentum projections). In the ground electronic and vibrational state (both symmetric), proton-exchange degeneracy imposes restrictions on the rotational levels, as the protons obey Fermi–Dirac statistics and the total molecular wavefunction must therefore be antisymmetric under proton exchange. Rotational wavefunctions with $K_{a} + K_{c}$ even are symmetric, whereas those with $K_{a} + K_{c}$ odd are antisymmetric. The nuclear-spin wavefunctions with total spin $1$ are symmetric (ortho), while those with total spin $0$ are antisymmetric (para). Consequently, ortho states correspond to rotational levels with $K_{a} + K_{c}$ odd, and para states to levels with $K_{a} + K_{c}$ even. The molecular ground state $0_{00}$ is a para state, and the lowest ortho state is $1_{01}$. Their energy difference is approximately $23.8$~$\mathrm{cm^{-1}}$ (or $\sim 3$~meV)~\cite{coudert1992}. Thus, at very low temperatures of $\sim 3$~K, as encountered both in astrophysical environments and in laboratory cooled gas jets, the population of an ensemble of water molecules will reside almost exclusively in the para ground state, assuming thermal equilibrium.

However, the situation changes when water undergoes the transition from the gas to the solid phase. The free rotational motion of the molecules becomes hindered and largely suppressed as intermolecular forces lock them into specific orientations within the crystal lattice. A fully general treatment of spin interactions in ice would require considering a large number of water molecules due to the high proton density in the lattice. In the approach proposed by Buntkowsky et al.~\cite{buntkowsky2008}, however, the problem is simplified by focusing on two nearest-neighbor H\textsubscript{2}O molecules, under the assumption that the rest of the lattice behaves in a similar manner. This reduces the system to four spins, i.e., the four proton spins, since the most abundant oxygen isotope has nuclear spin zero. 

\begin{figure}[h!]
	\centering
	\includegraphics[width=0.65\columnwidth]{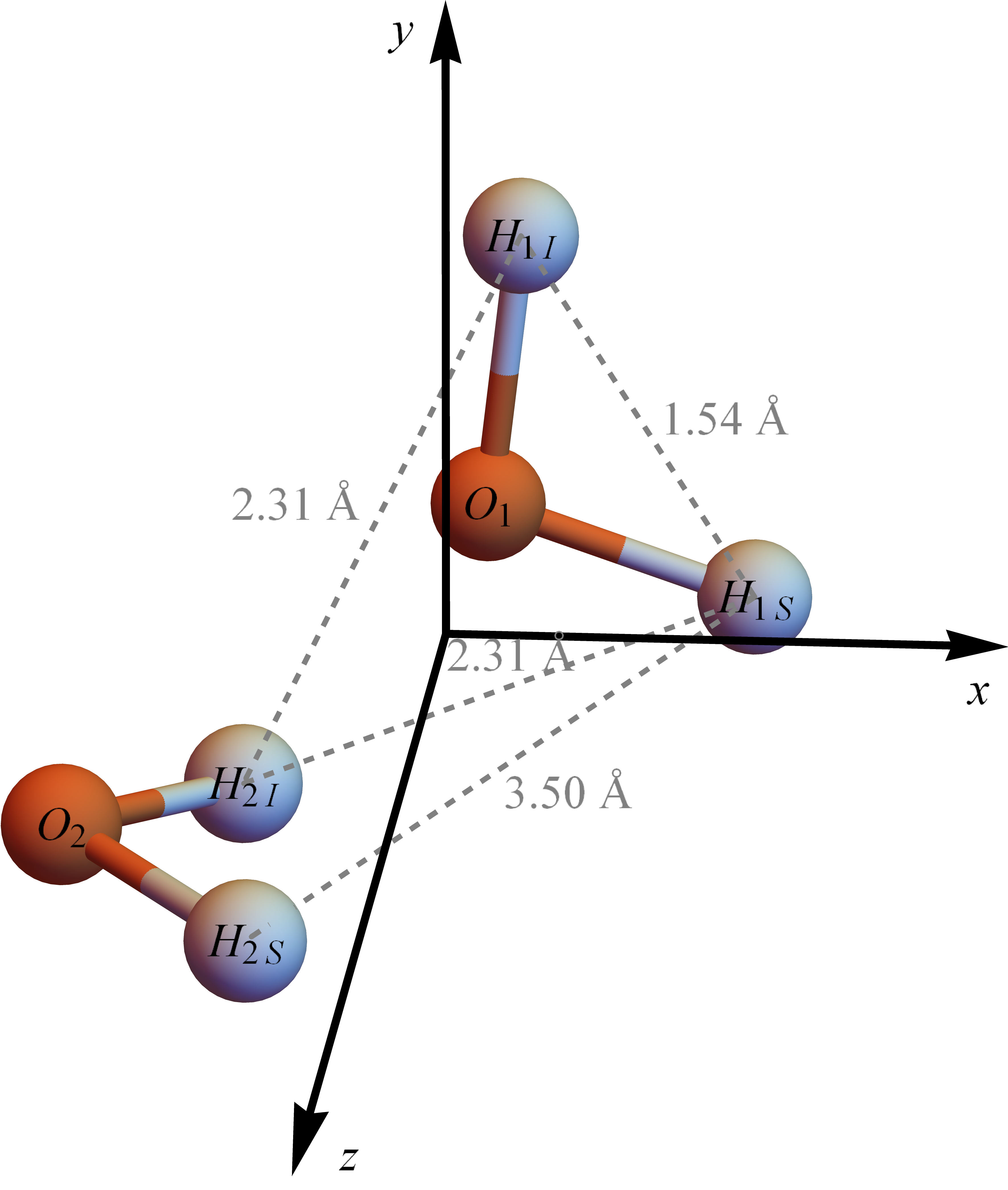}
	\caption{Distances between the four protons of two nearest-neighbor water molecules in the H\textsubscript{2}O ice lattice. The proton positions are adopted from Ref.~\cite{buntkowsky2025}, based on the X-ray structure of ice.}\label{figgeom}
\end{figure}

Figure~\ref{figgeom} illustrates the distances between the four protons, whose positions are taken from Ref.~\cite{buntkowsky2025} based on the X-ray structure of H\textsubscript{2}O ice. The two neighboring molecules are labeled with indices $1$ and $2$, and the two protons on each molecule are denoted by \rev{$S$ and $I$, yielding the four proton labels $H_{1 S}$, $H_{1 I}$, $H_{2 S}$, and $H_{2 I}$.} The corresponding spin operators are denoted by $\mathbf{S_{1}}$, $\mathbf{I_{1}}$, $\mathbf{S_{2}}$, and $\mathbf{I_{2}}$.

The interactions involving the proton spins that are relevant within this model, in the absence of external fields, are the magnetic dipole--dipole interactions and the indirect spin-spin interactions. These lead to the Hamiltonian~\cite{buntkowsky2008}:\rev{
\begin{equation}\label{eq:ham0}
	\begin{split}
		H_{0}  = & \mathbf{S_{1}} \cdot \mathbf{D_{1 S 1 I}} \cdot \mathbf{I_{1}} + \mathbf{S_{2}} \cdot \mathbf{D_{2 S 2 I}} \cdot \mathbf{I_{2}} + \mathbf{S_{1}} \cdot \mathbf{D_{1 S 2 S}} \cdot \mathbf{S_{2}} + \mathbf{S_{1}} \cdot \mathbf{D_{1 S 2 I}} \cdot \mathbf{I_{2}} \\
		& + \mathbf{I_{1}} \cdot \mathbf{D_{1 I 2 S}} \cdot \mathbf{S_{2}} +\mathbf{I_{1}} \cdot \mathbf{D_{1 I 2 I}} \cdot \mathbf{I_{2}} + \frac{J_{H_{2}O}}{\hbar^2} (\mathbf{S_{1}} \cdot \mathbf{I_{1}} + \mathbf{S_{2}} \cdot \mathbf{I_{2}}) ,
	\end{split}
\end{equation}}where the first two terms describe the dipolar interaction within the molecules $1$ and $2$, the next four terms represent the dipolar interaction between the proton spins of the two neighboring molecules, and the last two terms account for the $J$-coupling within each molecule. 

The dipolar interaction tensor $\mathbf{D_{K L}}$ is a rank-$2$ tensor, meaning it can be represented by a matrix with components $(D_{K L})_{i j}$, expressed in Cartesian coordinates as\rev{~\cite{abragam1961}}
\begin{equation}\label{eq:dip}
	(D_{K L})_{i j} = \frac{\mu_{0}}{4\pi} \bigg(\frac{g_{p} \mu_{N}}{\hbar}\bigg)^{2} \, \frac{1}{r_{K L} ^{5}} \big( r_{K L} ^{2} \delta_{i j} - 3 (r_{K L})_{i} (r_{K L})_{j} \big)
\end{equation}
with $i,j = x,y,z$, $r_{K L} = \left\vert \mathbf{r_{K L}}\right\vert$ the distance between the positions of the magnetic moments, and $(r_{K L})_{i}$ denoting the $i$-th Cartesian component of the separation vector $\mathbf{r_{K L}}$. For completeness, the magnetic dipole moment of a proton with spin $\mathbf{K}$ is $\boldsymbol{\mu_{K}} = \frac{g_{p} \mu_{N}}{\hbar} \, \mathbf{K}$, where $g_{p} = 5.586$ is the proton $g$ factor, and $\mu_{N} = 5.05 \times 10^{-27} \,\text{J/T}$ is the nuclear magneton.

The coupling constant $J_{H_{2}O}$ for a single water molecule in the ice lattice has not been reliably determined in the literature. It is, however, expected to be several orders of magnitude smaller than the dipolar couplings and can hence be neglected~\cite{buntkowsky2008, buntkowsky2006, buntkowsky1997}, $J_{H_{2}O} \approx 0$. Additional interactions of electric origin, such as those associated with molecular electric dipoles and hydrogen bonding, are not explicitly included in this work. Consequently, the employed Hamiltonian for the minimal model in the absence of external fields is purely magnetic and accounts solely for the dipole--dipole interactions between the four proton spins.   

To evaluate the spin dynamics governed by the Hamiltonian in Eq.~(\ref{eq:ham0}), it is advantageous to adopt representations in which the relevant operators take a simple form and the observables of interest can be computed efficiently. For a particle with spin $\mathbf{S}$, the spin states are commonly denoted by $\left \vert S \, m_{S} \right \rangle$, where $m_{S}$ is the spin projection along the quantization axis. Since the minimal model involves only protons, we omit the explicit spin quantum number $S$ and denote the states solely by their projections, $\left \vert m_{S} \right \rangle$. The spin state of the four-proton system is then constructed as the tensor product
\begin{equation}
	\left \vert m_{S_{1}} , m_{I_{1}} , m_{S_{2}} , m_{I_{2}} \right \rangle = \left \vert m_{S_{1}} \right \rangle \otimes \left \vert m_{I_{1}} \right \rangle \otimes \left \vert m_{S_{2}} \right \rangle \otimes \left \vert m_{I_{2}} \right \rangle .
\end{equation}
This set of states constitutes the uncoupled basis.

The coupled basis provides an alternative representation, whose usefulness will become evident below. Strictly speaking, this terminology is more appropriate for each individual molecule rather than for the pair of neighboring molecules as a whole; nevertheless, it provides a convenient description of the relevant spin symmetries. For molecules $1$ and $2$ the total nuclear spins $\mathbf{K_{1}}$ and $\mathbf{K_{2}}$ are defined as the addition of the two proton spins, i.e.,
\begin{eqnarray}
	\mathbf{K_{1}} & = & \mathbf{S_1} \otimes \mathds{1} + \mathds{1}\otimes\mathbf{I_1} \\
	\mathbf{K_{2}} & = & \mathbf{S_2} \otimes \mathds{1} + \mathds{1}\otimes\mathbf{I_2} ,
\end{eqnarray}
where $\mathds{1}$ denotes the identity operator in the corresponding single-molecule spin space. The eigenvalues of the $\mathbf{K_{1 , 2}}$ spin projections along the quantization axis are denoted by $ m_{K_{1 , 2}}$ (in units of $\hbar$). For each molecule, the spin states have the common form $\left \vert K_{i} \, m_{K_{i}} \right \rangle ,$ $i = 1, 2$; with ortho states defined by $K_{i} = 1$ and para states by $K_{i} = 0$. The spin state of the two-molecule system in this representation is written as
\begin{equation}
	\left \vert K_{1} \, m_{K_{1}} , K_{2} \, m_{K_{2}} \right \rangle = \left \vert K_{1} \, m_{K_{1}} \right \rangle \otimes \left \vert K_{2} \, m_{K_{2}} \right \rangle .
\end{equation}

The coupled and uncoupled bases are related by a unitary transformation,
\begin{align}
	\left \vert  K_{1} \, m_{K_{1}} , K_{2} \, m_{K_{2}} \right \rangle = & \sum_{m_{S_{1}} , m_{I_{1}}}  \sum_{m_{S_{2}} , m_{I_{2}}} \left\langle S_{1} \, m_{S_{1}} , I_{1} \, m_{I_{1}} \vert  K_{1} \, m_{K_{1}} \right \rangle \nonumber \\
	& \times \left\langle S_{2} \, m_{S_{2}} , I_{2} \, m_{I_{2}} \vert  K_{2} \, m_{K_{2}} \right \rangle 	\left \vert m_{S_{1}} , m_{I_{1}} , m_{S_{2}} , m_{I_{2}} \right \rangle , \label{eq:transf}
\end{align}
where the transformation elements $\left\langle S_{1} \, m_{S_{1}} , I_{1} \, m_{I_{1}} \vert  K_{1} \, m_{K_{1}} \right \rangle$ and $\left\langle S_{2} \, m_{S_{2}} , I_{2} \, m_{I_{2}} \vert  K_{2} \, m_{K_{2}} \right \rangle$ are Clebsch-Gordan coefficients, taken to be real by convention. As a result, the transformation  matrix between the two bases can be chosen orthogonal. The complete set of vector states composing the two representations along with the transformation matrix are provided in~\ref{app1}.

The Hamiltonian matrix can be directly evaluated in the uncoupled representation, as follows from Eq.~(\ref{eq:ham0}). The individual contributions from the dipolar interactions are obtained using Eq.~(\ref{eq:dip}). The complete H\textsubscript{2}O ice lattice is modeled as an ensemble of a large number of identical subsystems, each described by the minimal model. Consequently, the state of the full lattice at any time can be represented as a statistical mixture of the quantum states of this subsystem. On this basis, the density operator formalism is employed to study the spin dynamics.

The time evolution of the density operator $\rho$ for a system governed by a Hamiltonian $H$ is described by the Liouville--von Neumann equation,
\begin{equation}\label{eq:vonN}
	i\hbar\frac{\partial \rho}{\partial t} = [H, \rho] .
\end{equation}
For a time-independent Hamiltonian $H$ (e.g., $H_{0}$ in Eq.~(\ref{eq:ham0})) and an initial state $\rho(0)$ at $t=0$, the solution reads
\begin{equation}\label{eq:vonN2}
	\rho(t) = e^{-i H t/\hbar} \rho (0) e^{i H t/\hbar} .
\end{equation}
Thus, the time evolution can be obtained once the eigenequation for the Hamiltonian $H$ is solved. Such a process for a $16 \times 16$ matrix is carried out numerically. 

The density operator contains all physically relevant information about the ensemble and enables the calculation of measurable quantities. For an observable $O$, the ensemble average (or simply average) of $O$ is given by
\begin{equation}\label{eq:exp}
	\left\langle O \right \rangle = \mathrm{Tr}(\rho O) . 
\end{equation}
Owing to the invariance of the trace under unitary transformations, $\mathrm{Tr}(\rho O)$ may be evaluated in any convenient representation.

The primary quantities of interest in this work are the ortho and para populations, as well as the average spin projections. Specifically, for the two-molecule system, the ortho population corresponds to the sum of the populations of all coupled states with $K_{1} = K_{2} = 1$, while the para population is given by the population of the coupled state with $K_{1} = K_{2} = 0$, namely $\left \vert 0 \, 0 , 0 \, 0 \right \rangle$. These populations can be obtained from the relevant diagonal elements of the density matrix $\rho$, when written in the coupled representation, denoted $\rho_{c}$. Alternatively, Eq.~(\ref{eq:exp}) may be used with $\rho_{c}$ together with the projection matrices $P_{oo, c}$, $P_{pp, c}$, and $P_{op, c}$, which yield the ortho, para, and residual populations, respectively. These matrices are defined in~\ref{app1}, and since their sum yields the identity matrix, the total population is always unity. 

The average spin projections along the three Cartesian directions $x$, $y$, and $z$, as defined by the minimal model (see Fig.~\ref{figgeom}), can be computed using the Pauli matrices. The quantization axis is chosen to coincide with the $z$ axis of this geometry. A set of spin projections is associated with a magnetization whose components are directly related to the spin components. In contrast to the ortho/para populations, the average spin projections are most conveniently evaluated in the uncoupled basis. 

Below, we define the average spin projection components for each of the four proton spins employing the corresponding spin components,
\begin{equation}
	\begin{aligned}
	\left\langle m_{1 A, j} \right\rangle = \frac{1}{\hbar} \left\langle S_{1, j} \right \rangle , \quad \left\langle m_{1 B, j} \right\rangle = \frac{1}{\hbar} \left\langle I_{1, j} \right \rangle , \quad \\
	\left\langle m_{2 A, j} \right\rangle = \frac{1}{\hbar} \left\langle S_{2, j} \right \rangle , \quad \text{and} \quad \left\langle m_{2 B, j} \right\rangle = \frac{1}{\hbar} \left\langle I_{2, j} \right \rangle
	\end{aligned}
\end{equation}
with $j = x, y, z$. Details on the spin matrices are provided in~\ref{app1}. Although they are readily obtained in the uncoupled representation, a basis transformation can be applied to express them in the same representation as $\rho$ in order to evaluate the desired averages using Eq.~(\ref{eq:exp}). The physically relevant quantity is the total spin component of all four protons. The average total spin projection along the $j$th axis is therefore
\begin{equation}
	\left\langle M_{j} \right\rangle = \left\langle m_{1 A, j} \right\rangle + \left\langle m_{1 B, j} \right\rangle + \left\langle m_{2 A, j} \right\rangle +  \left\langle m_{2 B, j} \right\rangle .
\end{equation} 
The vector polarization along a given direction is defined as the corresponding average total spin projection normalized to its maximum value, $2$ (from $4$ proton-spin projections), i.e., $\left\langle M_{j} \right\rangle /2$.

An additional effect incorporated in the minimal model of Buntkowsky et al.~\cite{buntkowsky2008} is phase relaxation, also referred to as spin--spin relaxation, characterized by the relaxation time $T_{2}$. In practice, phase relaxation can be implemented by multiplying the off-diagonal elements of the density matrix by the factor $e^{-t/T_{2}}$. This approach is meaningful only when the density matrix is expressed in the eigenbasis of the Hamiltonian $H$. Equivalently, in that basis, one may introduce the matrix $\rho_{_{\Gamma}}$, whose diagonal elements vanish, while its off-diagonal elements are given by the corresponding elements of $\rho$ multiplied by $-\frac{i\hbar}{T_{2}}$. Therefore, the equation governing the spin dynamics, Eq.~(\ref{eq:vonN}), can be modified to include phase relaxation in the form
\begin{equation}\label{eq:vonN3}
	i\hbar\frac{\partial \rho}{\partial t} = [H, \rho] + \rho_{_{\Gamma}} \quad \text{with} \quad 
	{\rho_{_{\Gamma}}}_{m n} = \begin{cases}
		0, &   m = n \\
		-\frac{i\hbar}{T_{2}} \rho_{mn}, & m \neq n\end{cases} ,
\end{equation}
which provides a phenomenological description of coherence loss over time. The relaxation time is assumed constant and set to $T_{2} = 5.1$~ms~\cite{buntkowsky2008}.

\section{Spin dynamics in a static homogeneous magnetic field}
\label{theoryb}

The minimal model is extended to include the interaction with an external magnetic field $\mathbf{B}$. The magnetic moment of each of the four protons contributes an interaction term of the form
\begin{equation}
	- (1-\sigma)\boldsymbol{\mu} \cdot \mathbf{B} ,
\end{equation}
where $\sigma$ is the magnetic shielding constant, in general may be anisotropic and described by a tensor. Since $\sigma$ is small (e.g., for H\textsubscript{2}O at 25~$^{\circ}$C it is on the order of $10^{-5}$~\cite{tiesinga2021,modig2002}) and the present work employs a simplified toy model, this correction is neglected. Possible anisotropies of the proton $g$ factor within the lattice are also disregarded, and all proton magnetic moments are treated as isotropic. This is justified as the proton $g$ factor is fundamentally isotropic, while environment-induced anisotropies are negligible compared with the dipolar interaction strengths relevant here. Consequently, the interaction with an applied magnetic field $\mathbf{B}$ is incorporated by adding to the Hamiltonian the term
\begin{equation}\label{eq:hamb}
	H_{B} = -\frac{g_{p} \mu_{N}}{\hbar} (\mathbf{S_{1}}\cdot\mathbf{B} + \mathbf{I_{1}}\cdot\mathbf{B} + \mathbf{S_{2}}\cdot\mathbf{B} + \mathbf{I_{2}}\cdot\mathbf{B}) ,
\end{equation}
so that the total Hamiltonian becomes $H = H_{0} + H_{B}$.

In the following, a static and spatially uniform magnetic field directed along the $z$ axis is considered. Its possible influence on the phase-relaxation mechanism is neglected, and the same relaxation time $T_{2}$ as in the field-free case is adopted, as a microscopic description of field-dependent coherence loss is not available. The phase relaxation is incorporated in a phenomenological manner as explained in Sec.~\ref{theorya}. With these assumptions, the total Hamiltonian remains time independent, and the spin dynamics can be obtained from the solution of Eq.~(\ref{eq:vonN3}) by diagonalizing $H$.

\begin{figure}[h!]
	\centering
	\includegraphics[width=1.0\columnwidth]{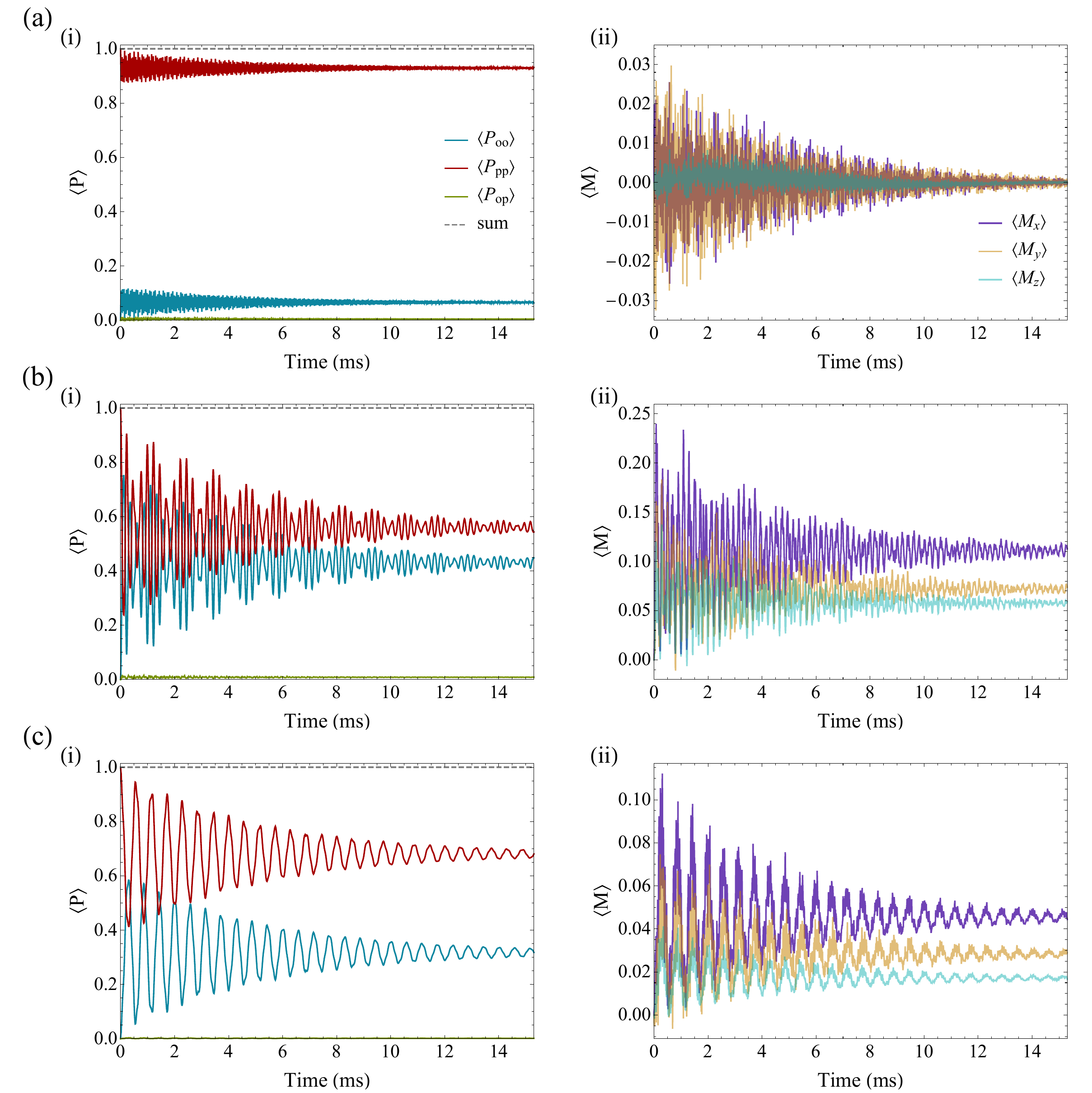}
	\caption{Time evolution of the populations $\left\langle P_{oo}\right\rangle$, $\left\langle P_{pp}\right\rangle$, and $\left\langle P_{op}\right\rangle$ along with the average total spin projection components $\left\langle M_{x} \right\rangle$, $\left\langle M_{y} \right\rangle$, and $\left\langle M_{z} \right\rangle$ for magnetic field values of $0$~mT (a), $1$~mT (b), and $2$~mT (c). All molecules are initially prepared in the para state.}\label{figpt}
\end{figure}

First, all H\textsubscript{2}O molecules are assumed to be in the para state\footnote{In the coupled representation this corresponds to a density matrix with unity in the last element ($16$th row and column) and zero elsewhere.} at $t = 0$, which is a valid description for the gas phase at very low temperatures $\sim 3$~K (see Sec.~\ref{theorya}). At this moment the molecules are assumed to enter the solid phase, and their subsequent spin dynamics are described within the minimal model. Figure~\ref{figpt} shows the time evolution of the ortho, para, and residual populations, together with the average total spin projections along the $x$, $y$, and $z$ directions, for magnetic field values of $0$, $1$, and $2$~mT during the first $15.3$~ms ($ = 3T_{2}$).

The oscillations between the ortho and para states arise because the eigenstates of a free rotor molecule (gas phase) and those in the rigid lattice are different, as explained in Ref.~\cite{buntkowsky2008}. This is expected as the Hamiltonian in the minimal model involves the dipolar interactions between the neighboring molecules, which occur once the lattice is formed, and are absent in the Hamiltonian of an isolated freely rotating molecule. As a result, even though the initial population resides entirely in a para state, a nonvanishing ortho--para conversion occurs \rev{since the system is initially found in an eigenstate of a free rotor and not in an eigenstate of the lattice minimal model}. This conversion is small in the absence of external fields (Fig.~\ref{figpt}a(i)) but can be manipulated and enhanced in the presence of static magnetic fields (Figs.~\ref{figpt}b,c(i)) \rev{as the magnetic-field coupling perturbs the lattice Hamiltonian, thereby modifying its eigenstates and eigenvalues and consequently the resulting spin dynamics.} The average total spin projections vanish when no field is applied (Fig.~\ref{figpt}a(ii)), whereas they acquire small values in the presence of an external field, corresponding to a vector polarization of a few percent along the three Cartesian directions. The phase relaxation, discussed in Sec.~\ref{theorya} and characterized by $T_{2}$, leads to a damping of the oscillation amplitudes, while the sum of populations remains unity at all times, since this process does not affect the diagonal elements of the density matrix.

\begin{figure}[h!]
	\centering
	\includegraphics[width=1.0\columnwidth]{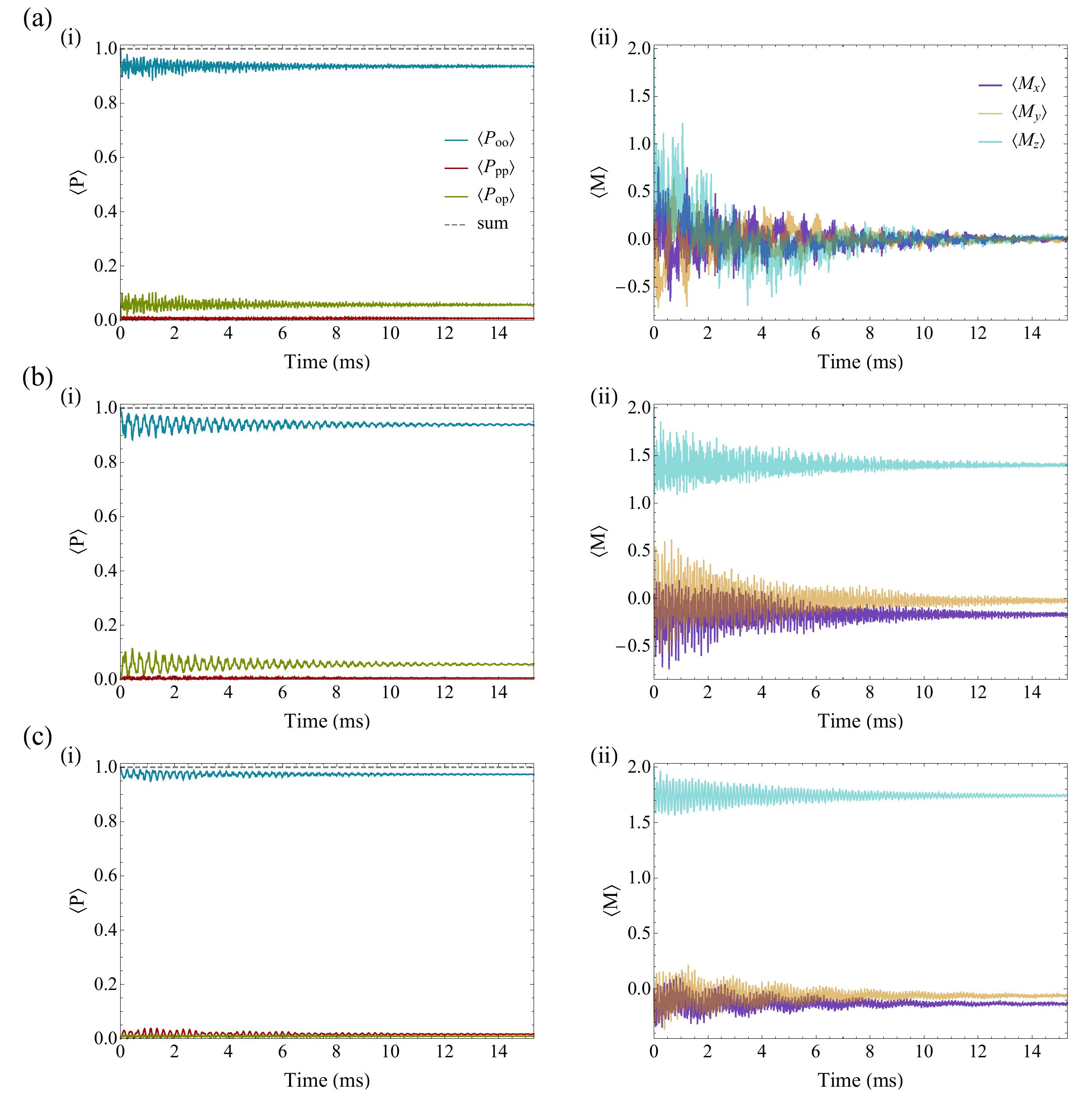}
	\caption{Time evolution of the populations $\left\langle P_{oo}\right\rangle$, $\left\langle P_{pp}\right\rangle$, and $\left\langle P_{op}\right\rangle$ along with the average total spin projection components $\left\langle M_{x} \right\rangle$, $\left\langle M_{y} \right\rangle$, and $\left\langle M_{z} \right\rangle$ for magnetic field values of $0$~mT (a), $0.5$~mT (b), and $1$~mT (c). The initial state corresponds to full spin polarization along the $z$ axis.}\label{figot}
\end{figure}

\rev{The previous example considered a rotationally cold ensemble with no initial nuclear-spin polarization.} We now consider the opposite limit, namely a beam that is fully spin polarized along the $z$ axis. Such polarization can be realized for the ortho species using optical techniques~\cite{sofikitis2015}, although the prepared molecules are then associated with nonzero rotational angular momentum (see Sec.~\ref{theorya}). Upon deposition into the solid phase, the free molecular rotation is quenched as the molecule becomes locked into a fixed orientation by the hydrogen-bond network. Consequently, the usual gas-phase ortho-para correlation between nuclear-spin symmetry and rotational states no longer applies, while the nuclear-spin symmetry itself is preserved. At $t=0$, all H\textsubscript{2}O molecules are therefore assumed to occupy an ortho state with their total nuclear spin aligned parallel to the $z$ axis\footnote{In the coupled representation this corresponds to a density matrix with unity in the first element (first row and column) and zero elsewhere.}.

Figure~\ref{figot} presents the time evolution of the ortho, para, and residual populations together with the average total spin projections along the $x$, $y$, and $z$ directions for magnetic fields of $0$, $0.5$, and $1$~mT during the first $15.3$~ms. As seen in Fig.~\ref{figot}a, the dipolar interaction alone not only drives ortho--para conversion but also leads to a rapid loss of the initial spin polarization on a sub-millisecond time scale. When a uniform magnetic field of the order of millitesla is applied parallel to the initial polarization, both the ortho--para conversion and the depolarization are suppressed. \rev{For sufficiently strong fields, the magnetic field interaction dominates the dipolar coupling and polarization losses become negligible.}

\begin{figure}[h!]
	\centering
	\includegraphics[width=1.0\columnwidth]{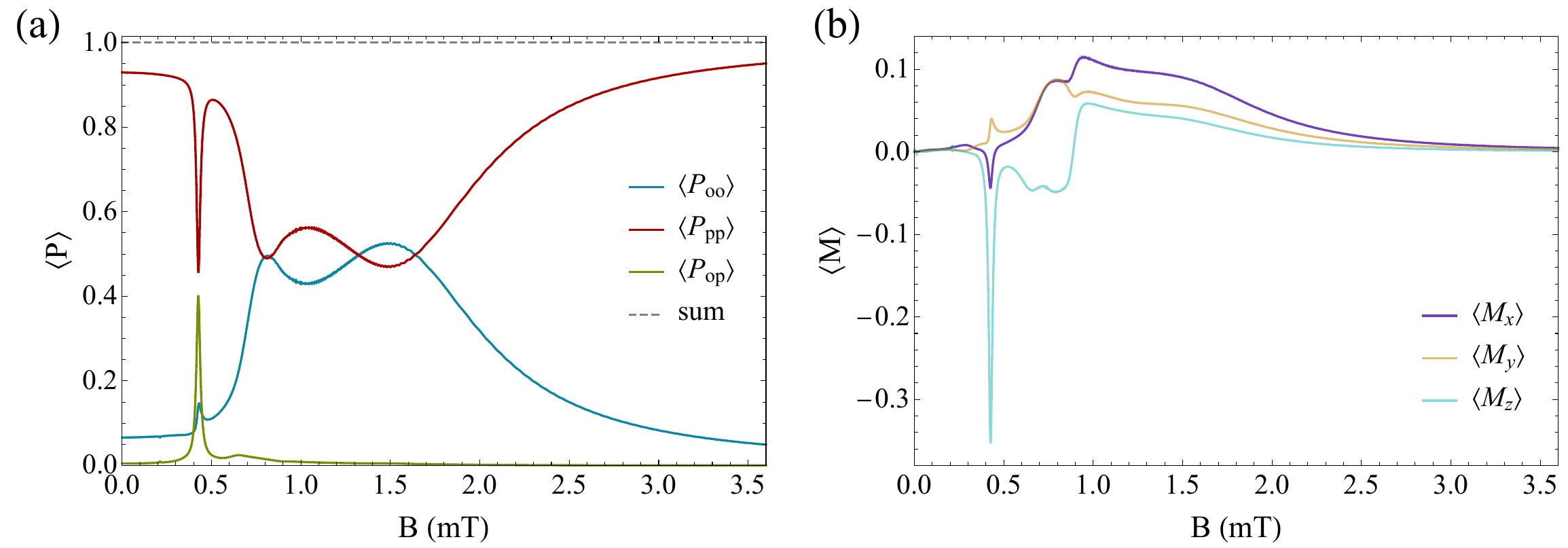}
	\caption{Populations $\left\langle P_{oo}\right\rangle$, $\left\langle P_{pp}\right\rangle$, and $\left\langle P_{op}\right\rangle$ (a) and average total spin projection components $\left\langle M_{x} \right\rangle$, $\left\langle M_{y} \right\rangle$, and $\left\langle M_{z} \right\rangle$ (b) at time $6 T_{2}$ as functions of the magnetic field. All molecules are initially prepared in the para state.}\label{figpb}
\end{figure}

To obtain a clearer \rev{picture} of how the two extreme initial states are influenced by the applied magnetic field, we evaluate the relevant populations and average total spin projections at time $6 T_{2}=30.6$~ms, when phase relaxation can be considered essentially complete, for a range of magnetic field strengths. Figures~\ref{figpb}a~and~\ref{figpb}b show these quantities assuming the para state is initially fully populated, for magnetic fields up to $3.6$~mT. For fields larger than approximately $7.6$~mT, the conversion of the para population falls below $1\%$, since the coupling to the magnetic field dominates over the dipolar coupling responsible for the ortho--para conversion. When the two interactions are \rev{comparable in strength}, namely for fields below about $2$~mT, more pronounced effects appear. In particular, the ortho population reaches a maximum of $52.6\%$ at $1.48$~mT, exceeding the para population. The para population exhibits a minimum of $45.5\%$ at $\sim 0.43$~mT. Interestingly, the para population at this field is transferred predominantly to the ortho-para ($op$) configuration rather than to the ortho-ortho ($oo$) configuration, with $\left\langle P_{op}\right\rangle = 40.2\%$. To further elucidate the spin dynamics at this field, the corresponding time evolution is provided in~\ref{app2} (see Fig.~\ref{figpbt}). Overall, the conversion of the para population for the various magnetic fields does not generally result in appreciable longitudinal or transverse polarization. The only \rev{exception} occurs at $\sim 0.43$~mT, where a longitudinal polarization of about $-17.7\%$ is obtained.

\begin{figure}[h!]
	\centering
	\includegraphics[width=1.0\columnwidth]{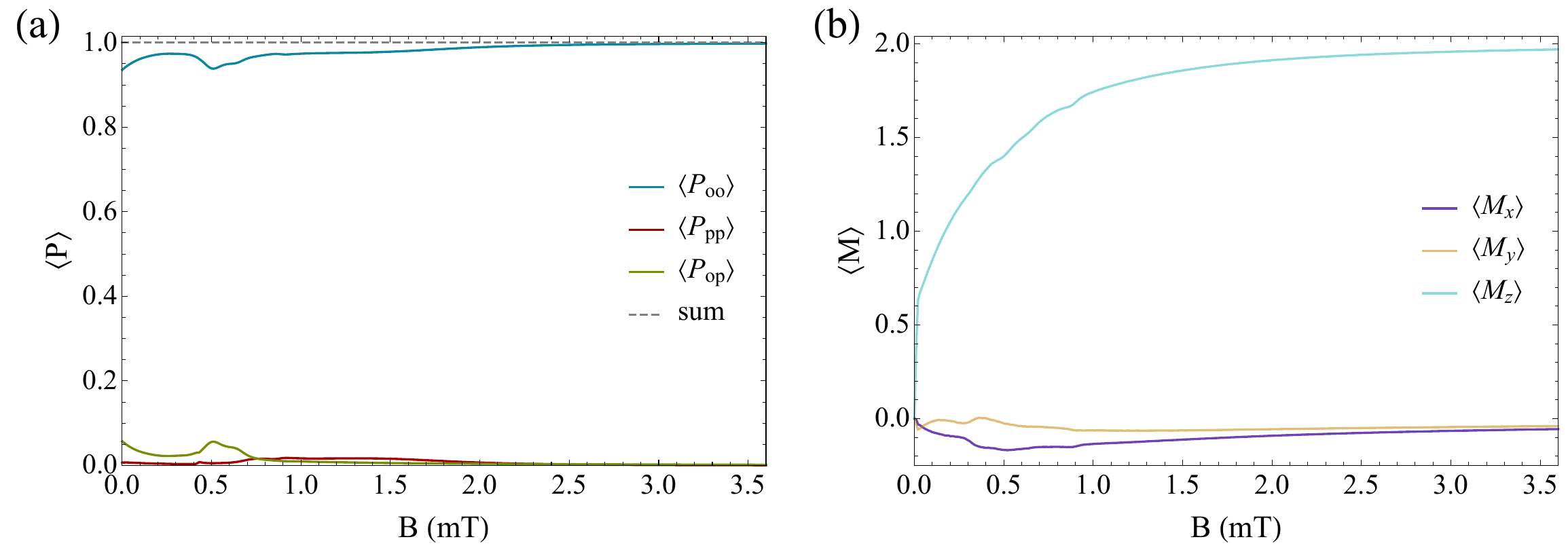}
	\caption{Populations $\left\langle P_{oo}\right\rangle$, $\left\langle P_{pp}\right\rangle$, and $\left\langle P_{op}\right\rangle$ (a) and average total spin projection components $\left\langle M_{x} \right\rangle$, $\left\langle M_{y} \right\rangle$, and $\left\langle M_{z} \right\rangle$ (b) at time $6 T_{2}$ as functions of the magnetic field. The initial state corresponds to full spin polarization along the $z$ axis.}\label{figob}
\end{figure}

Figure~\ref{figob} presents the same quantities as Fig.~\ref{figpb}, but for an initially fully polarized ensemble along the $z$ axis. The conversion of ortho population---predominantly to $op$ rather than to $pp$---remains small even at low magnetic fields. In contrast, the loss of longitudinal polarization is significant for fields below $\sim 1$~mT. For magnetic fields larger than $\sim 4.5$~mT, the longitudinal polarization loss decreases to below $1\%$, reflecting the dominance of the magnetic field coupling over the dipolar interactions.

\begin{figure}[h!]\rev{
		\centering
		\includegraphics[width=1.0\columnwidth]{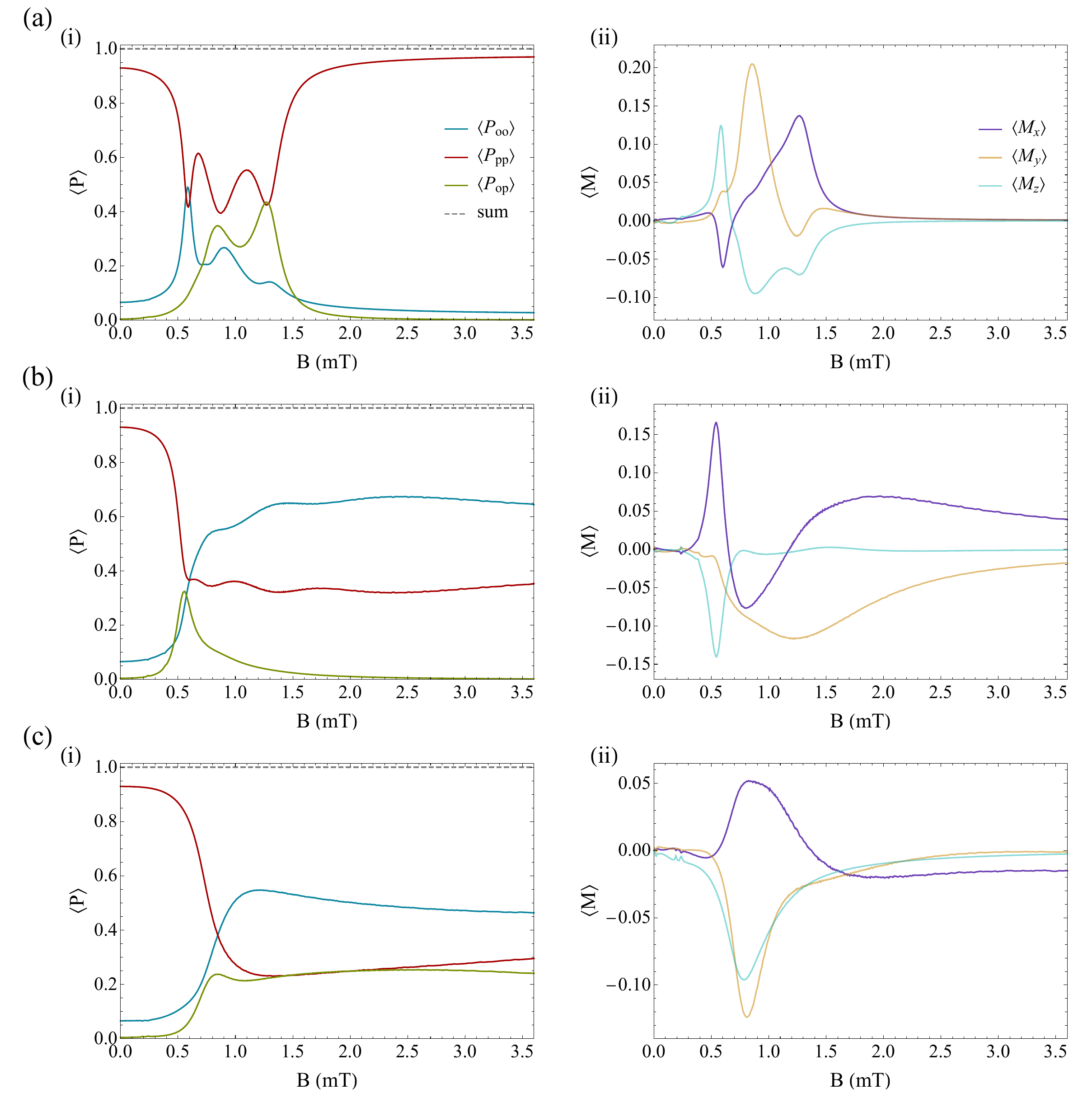}
		\caption{Populations $\left\langle P_{oo}\right\rangle$, $\left\langle P_{pp}\right\rangle$, and $\left\langle P_{op}\right\rangle$ along with the average total spin projection components $\left\langle M_{x} \right\rangle$, $\left\langle M_{y} \right\rangle$, and $\left\langle M_{z} \right\rangle$ at time $6 T_{2}$ as functions of the magnetic field. The molecular geometry is rotated by $45^{\circ}$ (a), $90^{\circ}$ (b), and $135^{\circ}$ (c) about the $x$ axis relative to the direction of the external magnetic field. All molecules are initially prepared in the para state.}\label{figprotx}}
\end{figure}

\rev{It should be noted that the above results are sensitive to the orientation of the two-molecule system relative to the external magnetic field. Since the dipolar interaction (see Eqs.~(\ref{eq:ham0}) and~(\ref{eq:dip})) depends explicitly on the spatial orientation of the internuclear vectors relative to the quantization axis, the resulting spin dynamics are expected to exhibit orientational dependence. Additional calculations were therefore performed by rotating the molecular geometry while keeping the magnetic field direction fixed. In other words, the Cartesian coordinate system shown in Fig.~\ref{figgeom} remains unchanged, whereas the proton coordinates are rotated. Figure~\ref{figprotx} illustrates how the results of Fig.~\ref{figpb} are modified when the molecular system is rotated by $45^{\circ}$, $90^{\circ}$, and $135^{\circ}$ about the $x$ axis for an initially pure para state. Corresponding results for rotations about the $y$ axis are presented in~\ref{app2} (see Fig.~\ref{figproty}). Depending on the orientation, the ortho--para population exchange can be either enhanced or suppressed at different magnetic fields; nevertheless, the most pronounced spin dynamics remain confined to the magnetic-field regime of a few mT. Specifically, the population exchange is suppressed at a lower magnetic field for a rotation of $45^{\circ}$ about the $x$ axis, whereas for rotations of $90^{\circ}$ and $135^{\circ}$ the ortho (para) population reaches approximately $58\%$ ($42\%$) and $51\%$ ($48\%$), respectively. This indicates that a substantial ortho population can persist at magnetic fields for which the field interaction dominates the spin dynamics. Variations are also observed in the average total spin projection components, although their qualitative behavior remains similar.}

\rev{For rotations about the $x$ or $y$ axis by an even multiple of $180^{\circ}$, the spin dynamics are identical to those of the unrotated system. Rotations by an odd multiple of $180^{\circ}$ likewise yield identical population dynamics, with the only difference being a reversal of the sign of the magnetization component along the rotation axis, while its magnitude remains unchanged. Rotations about the $z$ axis do not affect the ortho or para populations, nor the component $\left\langle M_{z} \right\rangle$. In contrast, $\left\langle M_{x} \right\rangle$ and $\left\langle M_{y} \right\rangle$ are modified because such rotations change the projection of $\left\langle \mathbf{M} \right\rangle$ on the fixed $x$ and $y$ axes defined in Fig.~\ref{figgeom}.}

\begin{figure}[h!]\rev{
		\centering
		\includegraphics[width=1.0\columnwidth]{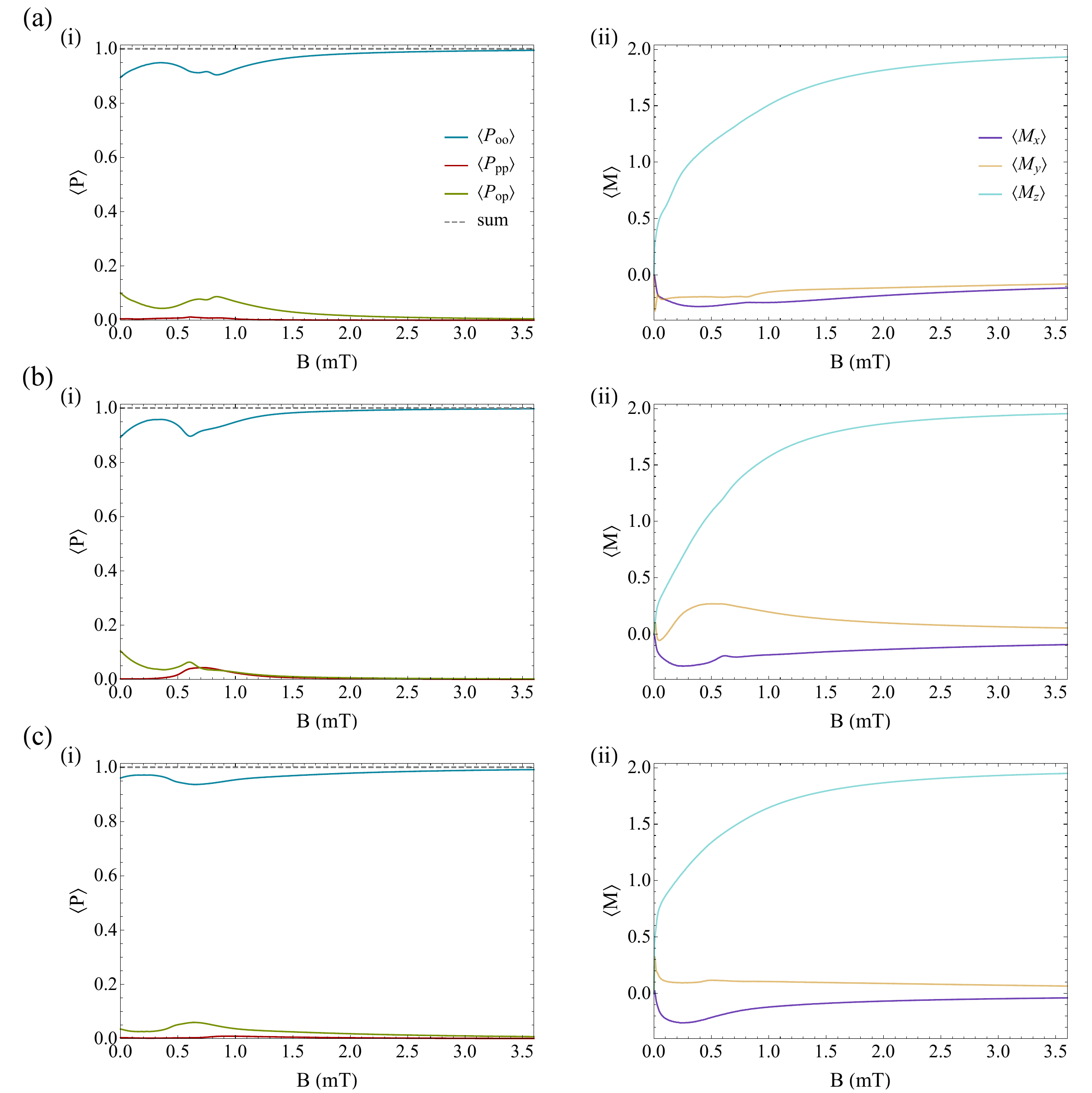}
		\caption{Populations $\left\langle P_{oo}\right\rangle$, $\left\langle P_{pp}\right\rangle$, and $\left\langle P_{op}\right\rangle$ along with the average total spin projection components $\left\langle M_{x} \right\rangle$, $\left\langle M_{y} \right\rangle$, and $\left\langle M_{z} \right\rangle$ at time $6 T_{2}$ as functions of the magnetic field. The molecular geometry is rotated by $45^{\circ}$ (a), $90^{\circ}$ (b), and $135^{\circ}$ (c) about the $x$ axis relative to the direction of the external magnetic field. The initial state corresponds to full spin polarization along the $z$ axis.}\label{figorthrotx}}
\end{figure}

\rev{However, the situation differs for initially spin-polarized molecules. The same rotations considered for the para case were also applied here, and the resulting spin dynamics are shown in Fig.~\ref{figorthrotx}. Corresponding results for rotations about the $y$ axis are provided in~\ref{app2} (Fig.~\ref{figorthroty}). In contrast to the para case, only a weak dependence on the molecular orientation is observed. For all examined orientations, the ortho--para conversion remains small even at low magnetic fields. Furthermore, for magnetic fields larger than $\sim 4.5$~mT, the loss of longitudinal polarization is effectively eliminated. This behavior reflects the dominance of the magnetic-field interaction over the dipolar interaction in this field regime.}

\rev{In summary, the spin dynamics in a static homogeneous magnetic field are governed by the interplay between dipolar interactions and magnetic field coupling. For weak magnetic fields ($0$--$2$~mT), the dipolar and magnetic-field interactions are of comparable strength, resulting in spin dynamics that depend on both the initial state and the relative orientation of the molecular geometry with respect to the external magnetic field. For magnetic fields larger than approximately $4$~mT, the magnetic field coupling dominates the spin dynamics and further increases in the magnetic-field strength have only a minor influence on the spin-state populations and magnetization. From a practical perspective, this implies that magnetic fields of a few millitesla are already sufficient to suppress dipolar depolarization during the gas-to-solid phase transition of nuclear-spin-polarized H\textsubscript{2}O molecules.}

\section{Spin dynamics in non-uniform/time-dependent magnetic fields}
\label{theoryc}

\rev{A static homogeneous magnetic field is commonly employed in coupled spin systems to perturb internal spin interactions and, at sufficiently high field strengths, effectively decouple otherwise interacting spin or angular momentum degrees of freedom. A well-known example is the decoupling of the rotational angular momentum and nuclear spins in molecular systems~\cite{wise2001, engels2015, kannis2018}. In many such systems, there is a characteristic magnetic field strength above which the coupled degrees of freedom are effectively decoupled, and the spin dynamics depend only weakly on further increases in the applied field. Likewise, the calculations presented in the previous section show that, during the gas-to-solid phase transition of H\textsubscript{2}O molecules, magnetic fields exceeding approximately $4$~mT result in spin dynamics that are only weakly affected by further increases in the magnetic-field strength.}

\rev{Exploiting the similarity between the spin dynamics of polarized particle beams and those of the minimal water-ice model introduced above, we proceed one step further and extend our analysis to include time-dependent magnetic field interactions. In particular, the application of a single sinusoidal magnetic-field pulse has been shown to effectively manipulate spins in atomic and molecular beams, leading to enhanced nuclear polarization~\cite{kannis2025}. This effect arises from magnetic-dipole transitions between spin states with unequal populations. For these transitions to be effective, the frequency and amplitude of the sinusoidal magnetic field must be appropriately tailored to the system under consideration. For simple systems such as ground-state H or D atoms, where only a single hyperfine interaction term is present, the excitation frequency is chosen to match the hyperfine splitting~\cite{kannis2025}. The magnetic field amplitude is subsequently scanned in order to determine the value required for a given application, such as maximizing the nuclear-spin polarization, depending on the initial conditions. However, the situation is more complicated for molecular systems due to the increased complexity of the effective hyperfine Hamiltonian. As a result, the optimal excitation frequency is determined by the dominant hyperfine interaction terms and may vary between different rotational levels. In a similar manner, for the minimal model considered here, the strengths of the intra- and intermolecular magnetic dipole--dipole interactions provide an estimate of the order of magnitude of the required frequency. Following the same procedure used for particle beams, we evaluate the relevant spin-dependent observables while scanning the magnetic-field amplitude.}

\begin{figure}[h!]
	\centering
	\includegraphics[width=1.0\columnwidth]{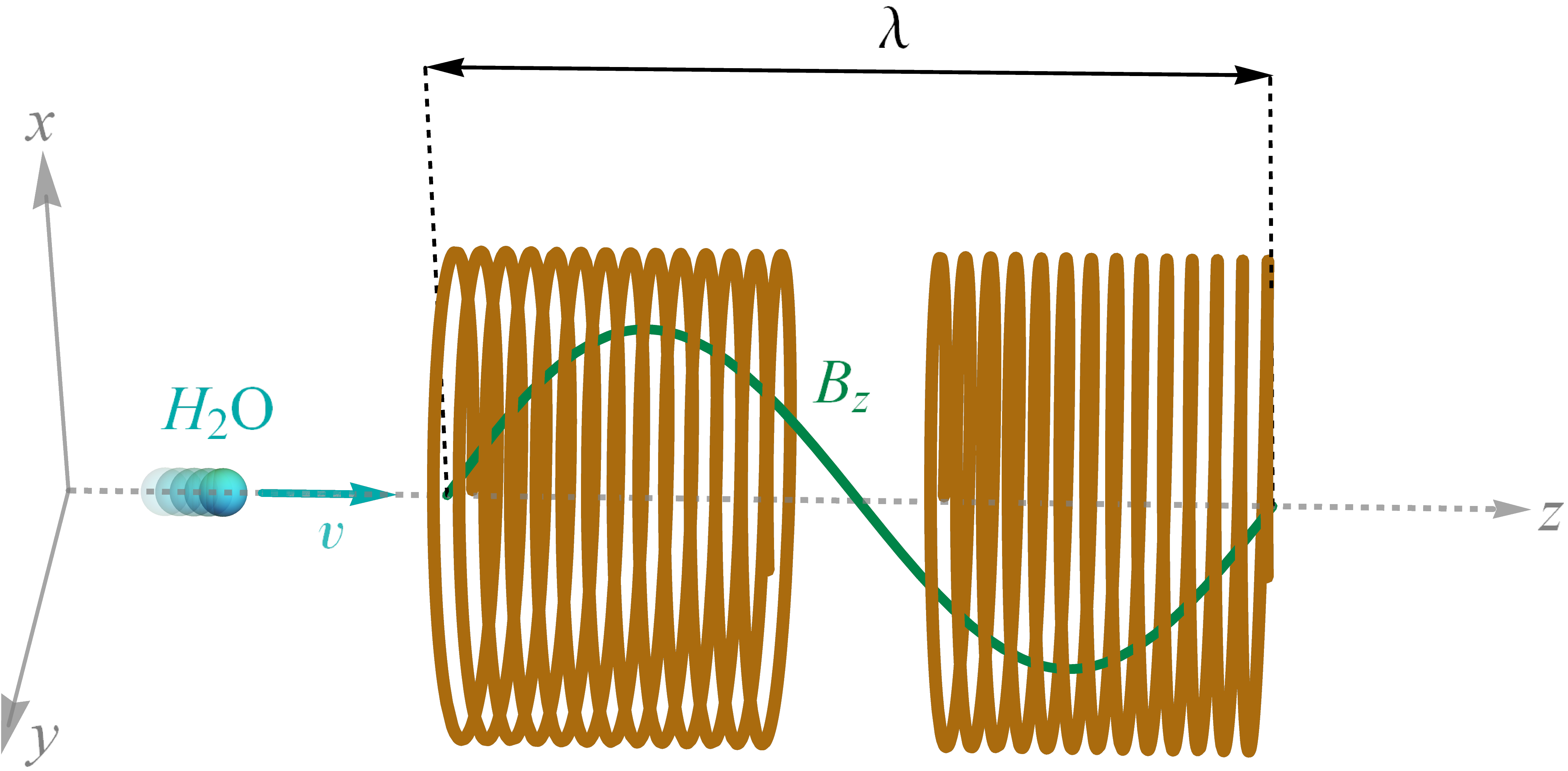}
	\caption{Schematic illustration of an H\textsubscript{2}O pellet moving with constant velocity $v$ through a longitudinal sinusoidal magnetic field. The wavelength $\lambda$ of the field is assumed to be much larger than the pellet radius $r_{p}$ ($\lambda\gg r_{p}$).}\label{figsetup}
\end{figure}

\rev{As a representative example, we consider a static, spatially sinusoidal magnetic field directed along the $z$ axis, where the latter is defined relative to the ice structure illustrated in Fig.~\ref{figgeom}.} Consider an H\textsubscript{2}O pellet formed at $t=0$, with all molecules initially in the para state, and moving with constant, nonrelativistic velocity $v$ parallel to the field axis. Although the magnetic field is static in the laboratory frame, the motion of the pellet renders the interaction effectively time dependent. A schematic illustration of this configuration is shown in Fig.~\ref{figsetup}.

For an axially symmetric coil configuration, with $r$ denoting the radial distance from the $z$ axis, the spatially varying magnetic field contains both longitudinal and radial components,
 \begin{equation}\label{eq:bfieldz}
	\mathbf{B} = B_{z} \hat{\mathbf{z}}  + B_{r} \hat{\mathbf{r}} = B_{0} \sin\Big(\frac{2 \pi z}{\lambda}\Big)  \hat{\mathbf{z}} - B_{0} \frac{\pi r}{\lambda}\cos\Big(\frac{2 \pi z}{\lambda}\Big) \hat{\mathbf{r}},
\end{equation}
where $B_{0}$ is the field amplitude and $\lambda$ its spatial period. In the following, we restrict ourselves to the regime $r\ll\lambda$, such that the longitudinal component $B_{z}$ dominates and the radial contribution can be neglected. In practice, this corresponds to a magnetic field configuration whose spatial extent is much larger than the size of the pellet.

The description above is given in the laboratory frame, where the pellet traverses a static sinusoidal field. An equivalent and often more convenient description is obtained in the pellet rest frame, where the pellet is stationary and the spatial variation of the field appears as an explicit time dependence. Substituting $z = v t$ into Eq.~(\ref{eq:bfieldz}) yields
\begin{equation}\label{eq:bfieldt}
	\mathbf{B} =  B_{0} \sin\Big(\frac{2 \pi v t}{\lambda}\Big)  \hat{\mathbf{z}} - B_{0} \frac{\pi r}{\lambda}\cos\Big(\frac{2 \pi v t}{\lambda}\Big) \hat{\mathbf{r}} .
\end{equation}

As discussed in Ref.~\cite{kannis2025} for related particle systems, the pellet rest frame provides a convenient framework for solving the spin dynamics. The translational motion is treated classically via a Galilean transformation between the laboratory and pellet frames, which is valid for nonrelativistic velocities. The spin degrees of freedom are treated quantum mechanically using the formalism introduced in the previous sections, now with a time-dependent magnetic field. The same formalism also applies to a static pellet subjected directly to a time-dependent magnetic field of the form given in Eq.~(\ref{eq:bfieldt}).

For concreteness, we assume a typical pellet velocity of $100$~$\text{m/s}$ (e.g., see Ref.~\cite{reistad2004}, where a hydrogen pellet target with $v \sim 90$~$\text{m/s}$ is reported). Considering a pellet size ($r_{p}$) on the order of millimeters, wavelengths on the order of meters are sufficient to ensure $r \ll \lambda$ for $0<r\leq r_{p}$, so that the longitudinal component dominates. Since it is convenient to express the interaction in terms of frequency, we define $f = v/\lambda$, which corresponds to the inverse of the time of flight $t_{f}$ across one spatial period. For $\lambda \sim 1$~m, this gives $f \sim 100$~Hz. In general, such a field can also either be directly applied to a static pellet or arise from a spatially varying magnetic field generated by coils moving relative to a stationary pellet. All of these scenarios are described by the interaction Hamiltonian of Eq.~(\ref{eq:hamb}) with $\mathbf{B} = B_{0} \sin (2 \pi f t)\hat{\mathbf{z}}$, assuming that only a single period of the field is applied, i.e., the magnetic field vanishes outside this time interval.

\begin{figure}[h!]
	\centering
	\includegraphics[width=1.0\columnwidth]{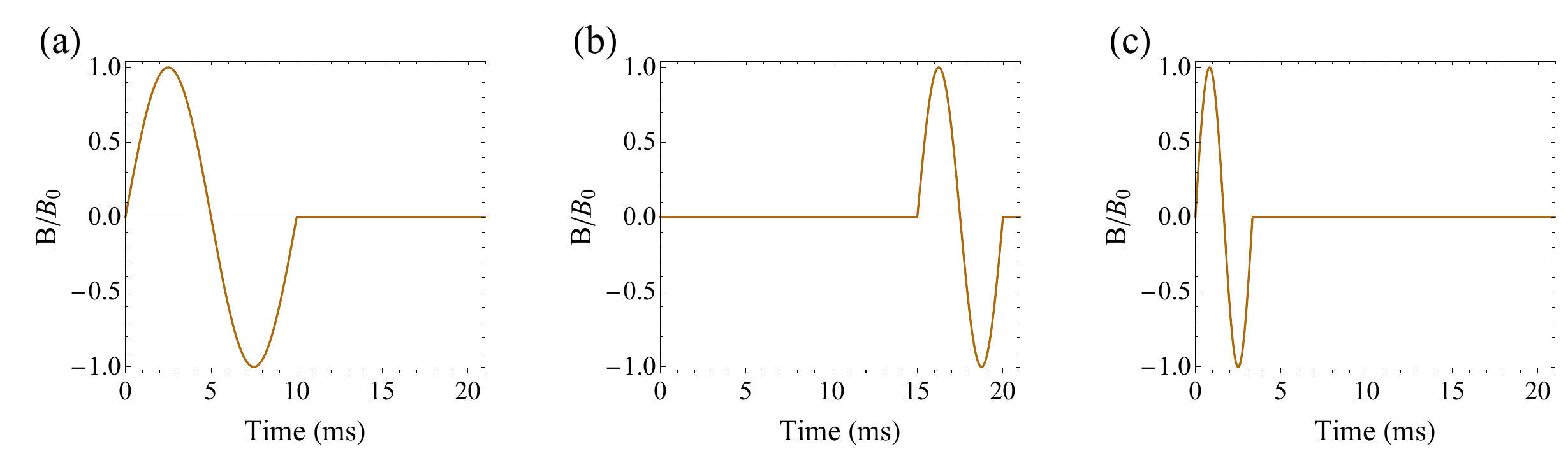}
	\caption{Magnetic field configurations in the pellet rest frame corresponding to frequencies $f = 100$~Hz (a), $f = 200$~Hz (b), and $f=300$~Hz (c). Pulses (a) and (c) are applied at $t=0$, whereas pulse (b) is applied at $t=15$~ms. The spin dynamics are studied for each configuration as a function of the magnetic field amplitude $B_{0}$.}\label{figpulses}
\end{figure}

Three representative magnetic field pulses, as experienced in the pellet rest frame, are considered. Figure~\ref{figpulses} shows their temporal profiles over the first $21$~ms. Their frequencies are $100$~Hz (a), $200$~Hz (b), and $300$~Hz (c). Pulses (a) and (c) are applied at $t=0$, while pulse (b) is applied at $t=15$~ms. The spin dynamics are investigated for these representative frequency--timing combinations by varying the magnetic-field amplitude $B_{0}$.

In contrast to the previous calculations, where either no field or static fields were considered, the total Hamiltonian $H_0 + H_B$ now contains explicit time dependence. Consequently, Eq.~(\ref{eq:vonN2}) is no longer applicable, and the von Neumann equation (Eq.~(\ref{eq:vonN})) must be solved numerically following the procedure described in Ref.~\cite{kannis2025}. This yields the time evolution of the density matrix $\rho$ in the absence of explicit phase relaxation. As discussed in earlier sections, phase relaxation is incorporated in a phenomenological manner, e.g., via Eq.~(\ref{eq:vonN3}). However, that treatment relies on a time-independent Hamiltonian and cannot be directly applied to a general time-dependent case, since the instantaneous eigensystem itself becomes time dependent. Therefore, the following procedure is adopted. At each time step, $\rho$ is obtained by solving Eq.~(\ref{eq:vonN}) in the uncoupled representation. The density matrix is then diagonalized, its off-diagonal elements are multiplied by a damping factor $e^{-t/T_{2}}$, and the matrix is transformed back to the uncoupled basis, where observables are evaluated. This approach effectively models the influence of decoherences while not accounting for possible modifications of the decoherence mechanism itself under magnetic field perturbations.

\begin{figure}[h!]
	\centering
	\includegraphics[width=1.0\columnwidth]{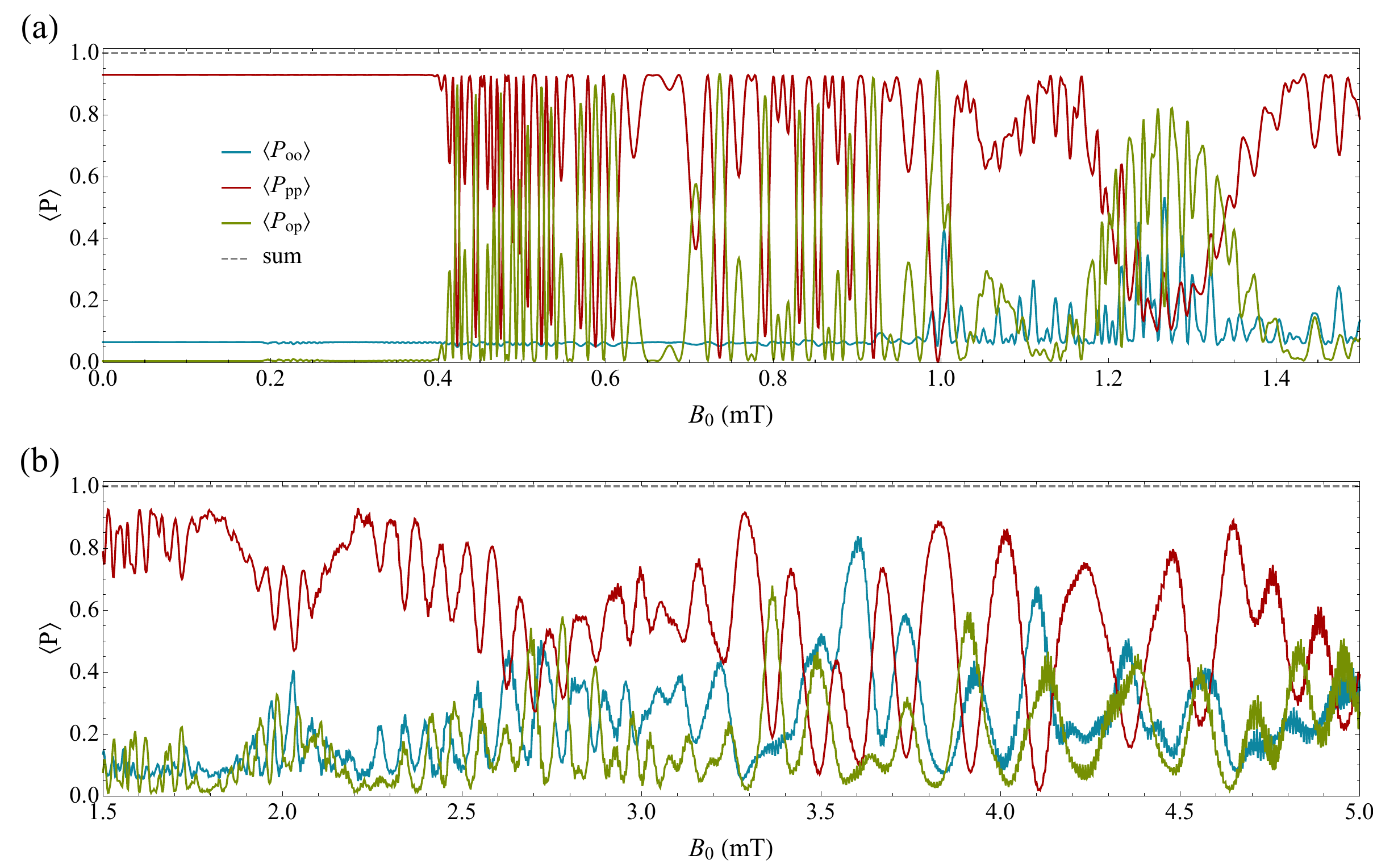}
	\caption{Populations $\left\langle P_{oo}\right\rangle$, $\left\langle P_{pp}\right\rangle$, and $\left\langle P_{op}\right\rangle$ at time $6 T_{2}$ as functions of the magnetic field amplitude $B_{0}$. All molecules are initially prepared in the para state and are subjected to a sinusoidal magnetic field with $f=100$~Hz applied at $t=0$ (see Fig.~\ref{figpulses}a).}\label{figp100hz}
\end{figure}

\begin{figure}[h!]
	\centering
	\includegraphics[width=1.0\columnwidth]{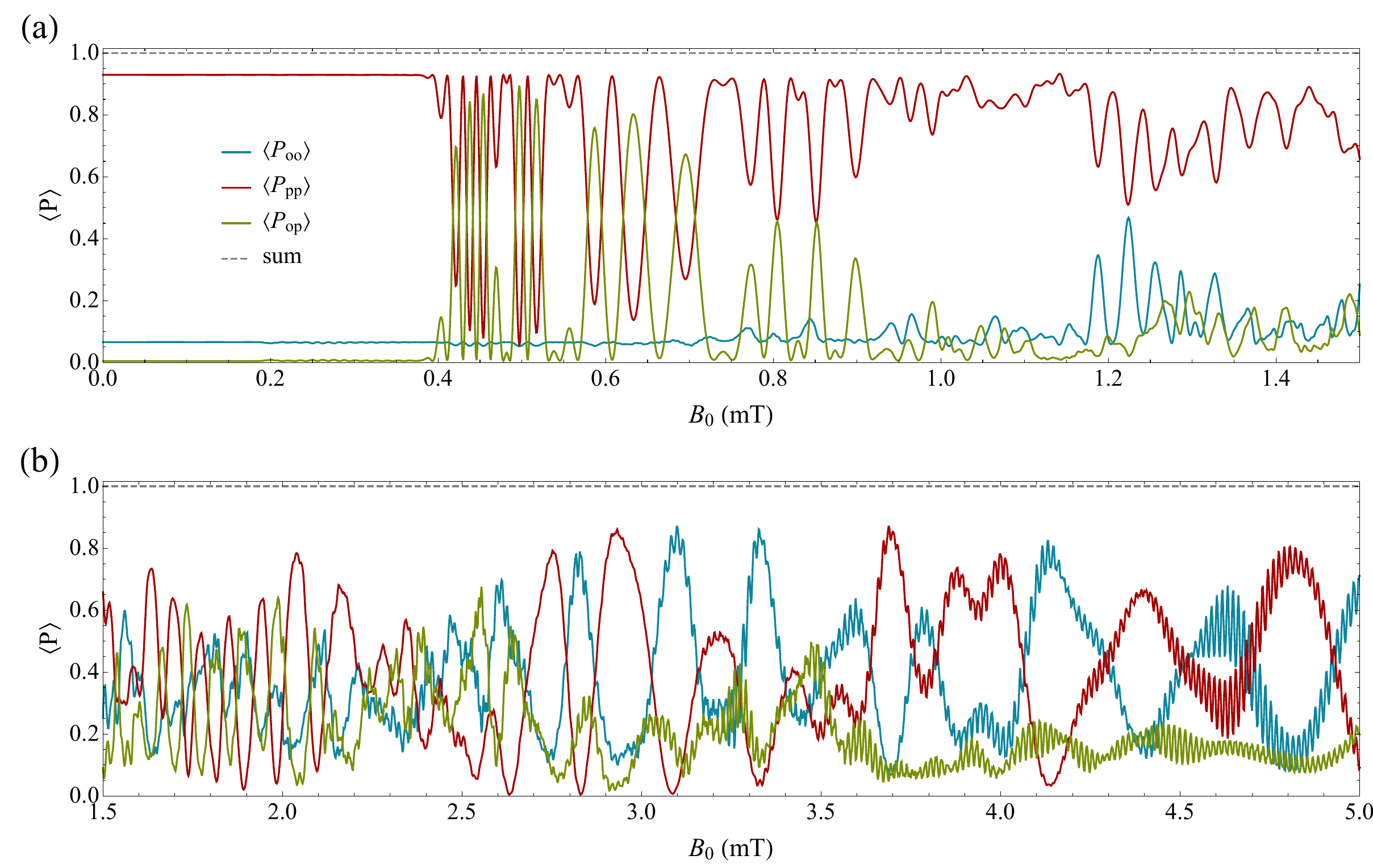}
	\caption{Populations $\left\langle P_{oo}\right\rangle$, $\left\langle P_{pp}\right\rangle$, and $\left\langle P_{op}\right\rangle$ at time $6 T_{2}$ as functions of the magnetic field amplitude $B_{0}$. All molecules are initially prepared in the para state and are subjected to a sinusoidal magnetic field with $f=200$~Hz at $t=15$~ms (see Fig.~\ref{figpulses}b).}\label{figp200hz}
\end{figure}

\begin{figure}[h!]
	\centering
	\includegraphics[width=1.0\columnwidth]{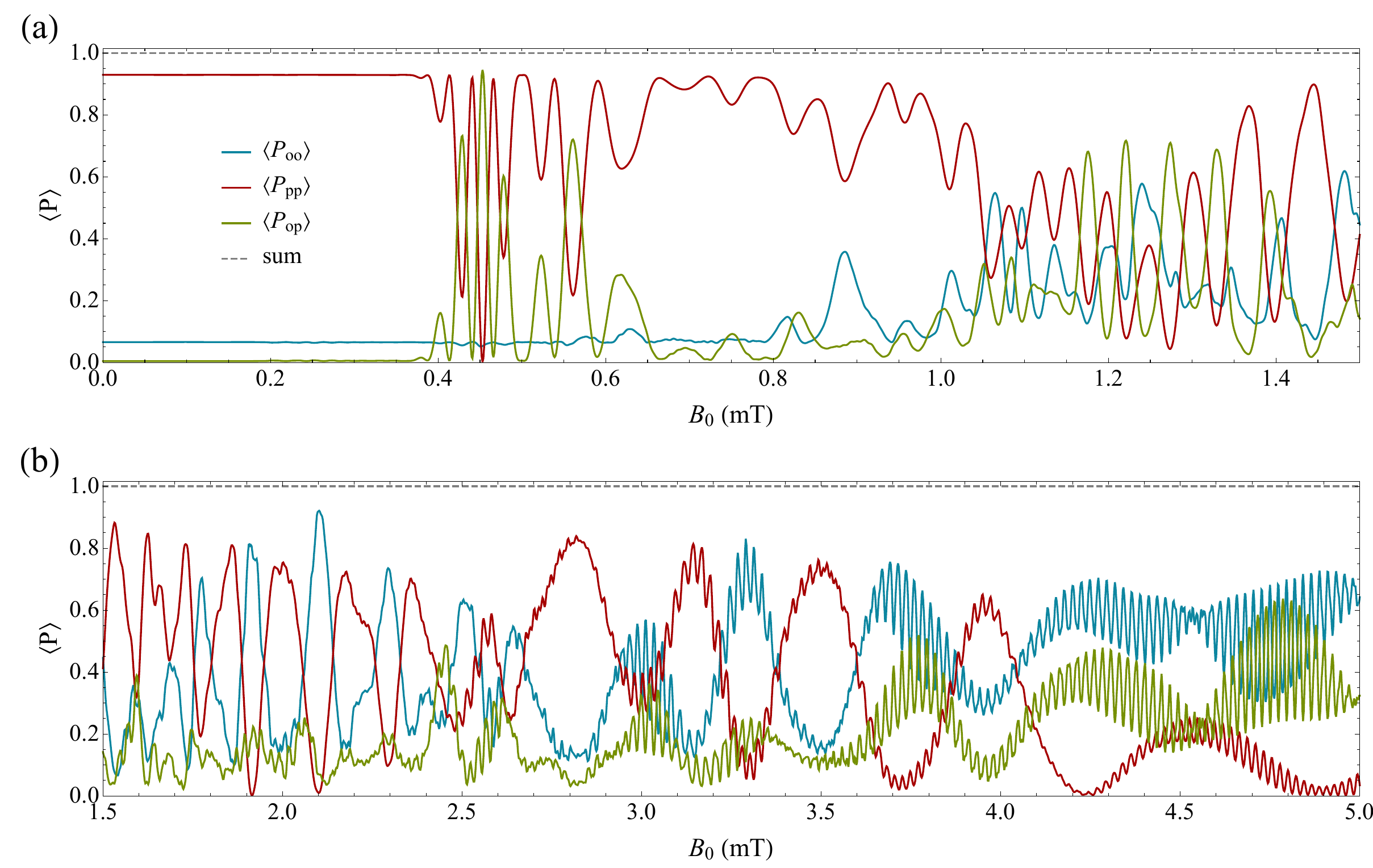}
	\caption{Populations $\left\langle P_{oo}\right\rangle$, $\left\langle P_{pp}\right\rangle$, and $\left\langle P_{op}\right\rangle$ at time $6 T_{2}$ as functions of the magnetic field amplitude $B_{0}$. All molecules are initially prepared in the para state and are subjected to a sinusoidal magnetic field with $f=300$~Hz at $t=0$ (see Fig.~\ref{figpulses}c).}\label{figp300hz}
\end{figure}

\begin{figure}[h!]
	\centering
	\includegraphics[width=1.0\columnwidth]{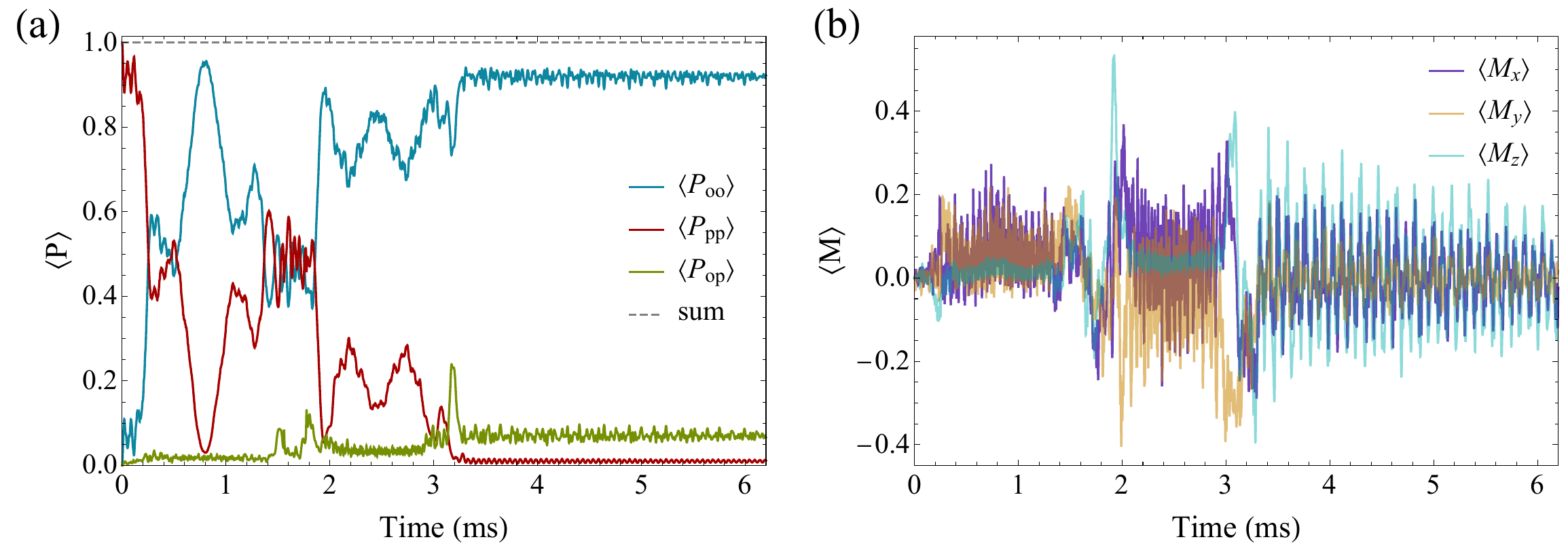}
	\caption{Time evolution of the populations $\left\langle P_{oo}\right\rangle$, $\left\langle P_{pp}\right\rangle$, and $\left\langle P_{op}\right\rangle$ (a), along with the average total spin projection components $\left\langle M_{x} \right\rangle$, $\left\langle M_{y} \right\rangle$, and $\left\langle M_{z} \right\rangle$ (b) for the magnetic field shown in Fig.~\ref{figpulses}c with $B_{0}= 2.1$~mT. All molecules are initially prepared in the para state.}\label{figt300Hz2p1mt}
\end{figure}

Figures~\ref{figp100hz}--\ref{figp300hz} present the spin-state populations at time $6 T_{2}$ for the magnetic field pulses depicted in Fig.~\ref{figpulses}, plotted as functions of the magnetic field amplitude $B_{0}$ in the range $0$--$5$~mT. The range is divided into two panels to resolve the rapid oscillations occurring at lower field strengths. The corresponding average total spin projections along the $x$, $y$, $z$ axes are given in~\ref{app2} (see Fig.~\ref{figpmabc}), as they do not exhibit appreciable spin polarization and are included for completeness. The populations, however, display pronounced \rev{field-dependent} structure. For $B_{0} \lesssim 1$~mT, rapid oscillations are observed, followed by slower oscillatory behavior as the field strength increases. In the higher-field region (panels (b)), the ortho population reaches maxima of $83.9\%$ (at $3.6$~mT), $86.8\%$ (at $3.1$~mT), and $91.9\%$ (at $2.1$~mT), respectively. To illustrate the last case in more detail, Fig.~\ref{figt300Hz2p1mt} shows the time evolution of the state populations during the pellet's passage through this field configuration. The ortho population stabilizes at $91.9\%$ by the end of the interaction region (after $3.3$~ms).

These results demonstrate that non-uniform magnetic fields in relative motion with H\textsubscript{2}O ice pellets can achieve substantially higher ortho--para conversion than static fields, which are typically limited to $\sim50\%$ (see Fig.~\ref{figpb}), when the molecules are initially in the para state. Further optimization of the magnetic field parameters could, in principle, enable even higher conversion efficiencies, as suggested by the nearly $92\%$ conversion obtained in the last of the examined cases. Moreover, as discussed in Ref.~\cite{kannis2025}, the sinusoidal-field technique is robust and can remain effective for a broad class of field shapes beyond purely sinusoidal waveforms. None of the configurations considered here, however, produces significant longitudinal or transverse spin polarization.

\section{Conclusion}
\label{conclusion}

The spin dynamics in H\textsubscript{2}O ice were studied using the simple two–nearest-neighbor molecule model (referred to as the minimal model) developed by Buntkowsky et al.~\cite{buntkowsky2008}, originally proposed as a mechanism for ortho--para conversion in ice. Restricting the interactions to magnetic dipolar couplings, we extended the model to include coupling to external magnetic fields. Both static homogeneous fields and spatially varying fields in relative motion with the ice pellet were considered. The analysis focused on the ortho-state populations and on the expectation values of the total spin projection components along the three Cartesian axes. All results were obtained numerically within the density operator formalism.

The extended model allowed us to study the interplay between dipolar couplings and magnetic-field interactions during the gas-to-solid phase transition. For static homogeneous fields and an initially para state, the maximum ortho population reached approximately $50\%$ at $1.48$~mT. A modest spin polarization ($\sim 18\%$) antiparallel to the $z$ axis was observed near $0.4$~mT. As the magnetic field strength increases beyond $\sim 4$~mT, the magnetic field coupling clearly dominates over the dipolar coupling, effectively suppressing ortho--para conversion. \rev{These results correspond to one particular orientation of the molecular geometry relative to the external magnetic field, for which the quantization axis is aligned with the $z$ axis of the molecular geometry. The calculations further demonstrate that the extent of ortho--para conversion depends on the orientation of the molecular geometry relative to the external magnetic field. While the quantitative conversion rates vary with orientation, the most pronounced spin dynamics remain confined to the magnetic-field regime of a few millitesla. Above approximately $4$~mT, the spin-state populations and magnetization components become largely insensitive to further increases in the magnetic field strength.} For an initially fully polarized ensemble, magnetic fields of approximately $4.5$~mT or larger are sufficient to effectively prevent loss of longitudinal polarization. \rev{This behavior is essentially independent of the molecular orientation, provided that the initial spin polarization is parallel to the external magnetic field. From a practical perspective, these results indicate that magnetic fields of only a few millitesla are sufficient to preserve the nuclear-spin polarization during the gas-to-solid phase transition within the framework of the present model.}

For spatially varying magnetic fields, scenarios involving relative motion between the field and an ice pellet were examined. In the rest frame of the ice, this corresponds to a time-dependent interaction, treated following Ref.~\cite{kannis2025}. The analysis was restricted to sinusoidal fields, chosen to overcome a limitation of static fields, namely the upper bound on the achievable ortho population from an initial exclusively para population. The considered configurations successfully exceeded this bound, yielding ortho populations as high as $\sim 80$–$90\%$. In general, rapid population oscillations were observed for field strengths below $\sim 1$~mT, while the dynamics slowed with increasing field strength. No significant polarization of the spin projection components was observed in these cases.

The modeling framework presented here is not restricted to water but can be extended to other systems, including proposals involving the freezing and collection of nuclear-spin-polarized beams~\cite{kannis2021,kannis20252}. By focusing on dominant nearest-neighbor spin interactions, the approach provides estimates of the external magnetic field strengths required to avoid depolarization of prepared spin states. For such applications, the model can be extended to incorporate additional interactions, such as those associated with the deuteron's electric quadrupole moment~\cite{abragam1961,buntkowsky20062}, enabling application to deuterated systems such as D\textsubscript{2}O, HDO, and D\textsubscript{2}. By contrast, H\textsubscript{2}~\cite{buntkowsky2006} can be treated within the present framework without further modification due to its analogous nuclear-spin structure.






\section*{Acknowledgments}
We gratefully acknowledge Prof.~Buntkowsky and his group for providing the material necessary to reproduce the minimal model proposed in Ref.~\cite{buntkowsky2008}, as well as for valuable discussions and continuous support. This work was funded by the Deutsche Forschungsgemeinschaft (DFG, German Research Foundation) under Grant No.~533904660.

\appendix
\section{Uncoupled and coupled spin states}
\label{app1}

Both the uncoupled and coupled bases contain $16$ kets. Adopting the conventions introduced in Sec.~\ref{theorya}, we list the basis elements below and specify the ordering used in the calculations.

\begin{align}
	\begin{array}{c}
		\text{uncoupled basis} \\[1pt]
		\{\left\vert m_{S_{1}} , m_{I_{1}} , m_{S_{2}} , m_{I_{2}} \right\rangle\}
	\end{array}
	\;:\;
	\Bigg\{
	& \left\vert  \frac{1}{2} , \frac{1}{2} , \frac{1}{2} , \frac{1}{2} \right\rangle , \,  \left\vert \frac{1}{2} , \frac{1}{2} , \frac{1}{2} , -\frac{1}{2} \right\rangle , \, \left\vert \frac{1}{2} , \frac{1}{2} , -\frac{1}{2} , \frac{1}{2} \right\rangle , \nonumber \\
	& \left\vert \frac{1}{2} , \frac{1}{2} , -\frac{1}{2} , -\frac{1}{2} \right\rangle , \, \left\vert \frac{1}{2}, -\frac{1}{2} , \frac{1}{2} , \frac{1}{2} \right\rangle , \, \left\vert \frac{1}{2}, -\frac{1}{2} , \frac{1}{2} , -\frac{1}{2} \right\rangle , \nonumber \\
	& \left\vert \frac{1}{2}, -\frac{1}{2} , -\frac{1}{2} , \frac{1}{2} \right\rangle , \, \left\vert \frac{1}{2}, -\frac{1}{2} , -\frac{1}{2} , -\frac{1}{2} \right\rangle , \, \left\vert -\frac{1}{2}, \frac{1}{2} , \frac{1}{2} , \frac{1}{2} \right\rangle , \nonumber \\
	& \left\vert -\frac{1}{2}, \frac{1}{2} , \frac{1}{2} , -\frac{1}{2} \right\rangle , \, \left\vert -\frac{1}{2}, \frac{1}{2} , -\frac{1}{2} , \frac{1}{2} \right\rangle , \, \left\vert -\frac{1}{2}, \frac{1}{2} , -\frac{1}{2} , - \frac{1}{2} \right\rangle , \nonumber \\
	& \left\vert -\frac{1}{2}, - \frac{1}{2} , \frac{1}{2} , \frac{1}{2} \right\rangle , \, \left\vert -\frac{1}{2}, - \frac{1}{2} , \frac{1}{2} , -\frac{1}{2} \right\rangle , \, \left\vert -\frac{1}{2}, - \frac{1}{2} , -\frac{1}{2} , \frac{1}{2} \right\rangle , \nonumber \\
	& \left\vert -\frac{1}{2}, - \frac{1}{2} , -\frac{1}{2} , - \frac{1}{2} \right\rangle
	\Bigg\}
	\\
	\begin{array}{c}
	\text{coupled basis} \\[1pt]
	\{\left\vert K_{1} \, m_{K_{1}} , K_{2} \, m_{K_{2}} \right\rangle\}
	\end{array}
	\;:\;
	\Big\{
	&\left\vert  1 \, 1 , 1 \, 1 \right\rangle , \,  \left\vert 1 \, 1 , 1 \, 0 \right\rangle , \, \left\vert 1 \, 1 , 1 \, -1 \right\rangle , \, \left\vert 1 \, 1 , 0 \, 0 \right\rangle , \nonumber \\
	& \left\vert 1 \, 0 , 1 \, 1 \right\rangle , \, \left\vert 1 \, 0 , 1 \, 0 \right\rangle , \, \left\vert 1 \, 0 , 1 \, -1 \right\rangle , \, \left\vert 1 \, 0 , 0 \, 0  \right\rangle , \nonumber \\
	& \left\vert 1 \, -1 , 1 \, 1 \right\rangle , \, \left\vert 1 \, -1 , 1 \, 0 \right\rangle , \, \left\vert 1 \, -1 , 1 \, -1 \right\rangle , \, \left\vert 1 \, -1 , 0 \, 0 \right\rangle ,  \nonumber \\
	& \left\vert 0 \, 0 , 1 \, 1 \right\rangle , \, \left\vert 0 \, 0 , 1 \, 0 \right\rangle , \, \left\vert 0 \, 0 , 1 \, -1 \right\rangle , \, \left\vert0 \, 0 , 0 \, 0 \right\rangle
\Big\}
\end{align}

Using Eq.~(\ref{eq:transf}) and the standard convention that Clebsch-Gordan coefficients are real, one obtains the following relations:
\begin{subequations}
	\begin{align}\label{eq:c1}
		&\left\vert 1 \, 1 , 1 \, 1 \right\rangle = \left\vert \frac{1}{2} , \frac{1}{2} , \frac{1}{2} , \frac{1}{2} \right\rangle  \\ \label{eq:c2}
		&\left\vert 1 \, 1 , 1 \, 0 \right\rangle = \frac{1}{\sqrt{2}} \Bigg( \left\vert \frac{1}{2} , \frac{1}{2} , \frac{1}{2} , -\frac{1}{2} \right\rangle + \left\vert \frac{1}{2} , \frac{1}{2} , -\frac{1}{2} , \frac{1}{2} \right\rangle \Bigg)  \\ \label{eq:c3}
		&\left\vert 1 \, 1 , 1 \, -1 \right\rangle = \left\vert \frac{1}{2} , \frac{1}{2} , -\frac{1}{2} , -\frac{1}{2} \right\rangle  \\ \label{eq:c4}
		&\left\vert 1 \, 1 , 0 \, 0 \right\rangle = \frac{1}{\sqrt{2}} \Bigg( \left\vert \frac{1}{2} , \frac{1}{2} , \frac{1}{2} , -\frac{1}{2} \right\rangle - \left\vert \frac{1}{2} , \frac{1}{2} , -\frac{1}{2} , \frac{1}{2} \right\rangle \Bigg)  \\ \label{eq:c5}
		&\left\vert 1 \, 0 , 1 \, 1 \right\rangle = \frac{1}{\sqrt{2}} \Bigg( \left\vert \frac{1}{2} , -\frac{1}{2} , \frac{1}{2} , \frac{1}{2} \right\rangle + \left\vert -\frac{1}{2} , \frac{1}{2} , \frac{1}{2} , \frac{1}{2} \right\rangle \Bigg)  \\  \label{eq:c6}
		&\left\vert 1 \, 0 , 1 \, 0 \right\rangle  =
		\begin{aligned}[t]
		 \frac{1}{2} \Bigg( &\left\vert \frac{1}{2} , -\frac{1}{2} , \frac{1}{2} , -\frac{1}{2} \right\rangle + \left\vert \frac{1}{2} , -\frac{1}{2} , -\frac{1}{2} , \frac{1}{2} \right\rangle \\
		+ &\left\vert -\frac{1}{2} , \frac{1}{2} , \frac{1}{2} , - \frac{1}{2} \right\rangle + \left\vert -\frac{1}{2} , \frac{1}{2} , -\frac{1}{2} , \frac{1}{2} \right\rangle \Bigg)
		\end{aligned}  
		\\ \label{eq:c7}
		&\left\vert 1 \, 0 , 1 \, -1 \right\rangle = \frac{1}{\sqrt{2}} \Bigg( \left\vert \frac{1}{2} , -\frac{1}{2} , -\frac{1}{2} , -\frac{1}{2} \right\rangle + \left\vert -\frac{1}{2} , \frac{1}{2} , -\frac{1}{2} , -\frac{1}{2} \right\rangle \Bigg)  \\ \label{eq:c8}
		&\left\vert 1 \, 0 , 0 \, 0 \right\rangle =
		\begin{aligned}[t]
		 \frac{1}{2} \Bigg( &\left\vert \frac{1}{2} , -\frac{1}{2} , \frac{1}{2} , -\frac{1}{2} \right\rangle - \left\vert \frac{1}{2} , -\frac{1}{2} , -\frac{1}{2} , \frac{1}{2} \right\rangle \\
		 + & \left\vert -\frac{1}{2} , \frac{1}{2} , \frac{1}{2} , - \frac{1}{2} \right\rangle - \left\vert -\frac{1}{2} , \frac{1}{2} , -\frac{1}{2} , \frac{1}{2} \right\rangle \Bigg) 
		 \end{aligned}
		 \\ \label{eq:c9}
		&\left\vert 1 \, -1 , 1 \, 1 \right\rangle = \left\vert -\frac{1}{2} , -\frac{1}{2} , \frac{1}{2} , \frac{1}{2} \right\rangle \\ \label{eq:c10}
		&\left\vert 1 \, -1 , 1 \, 0 \right\rangle = \frac{1}{\sqrt{2}} \Bigg( \left\vert -\frac{1}{2} , -\frac{1}{2} , \frac{1}{2} , -\frac{1}{2} \right\rangle + \left\vert -\frac{1}{2} , -\frac{1}{2} , -\frac{1}{2} , \frac{1}{2} \right\rangle \Bigg) \\ \label{eq:c11}
		&\left\vert 1 \, -1 , 1 \, -1 \right\rangle = \left\vert -\frac{1}{2} , -\frac{1}{2} , -\frac{1}{2} , -\frac{1}{2} \right\rangle \\ \label{eq:c12}
		&\left\vert 1 \, -1 , 0 \, 0 \right\rangle = \frac{1}{\sqrt{2}} \Bigg( \left\vert -\frac{1}{2} , -\frac{1}{2} , \frac{1}{2} , -\frac{1}{2} \right\rangle - \left\vert -\frac{1}{2} , -\frac{1}{2} , -\frac{1}{2} , \frac{1}{2} \right\rangle \Bigg) \\ \label{eq:c13}
		&\left\vert 0 \, 0 , 1 \, 1 \right\rangle = \frac{1}{\sqrt{2}} \Bigg( \left\vert \frac{1}{2} , -\frac{1}{2} , \frac{1}{2} , \frac{1}{2} \right\rangle - \left\vert -\frac{1}{2} , \frac{1}{2} , \frac{1}{2} , \frac{1}{2} \right\rangle \Bigg) \\ \label{eq:c14}
		&\left\vert 0 \, 0 , 1 \, 0 \right\rangle =
		\begin{aligned}[t]
		 \frac{1}{2} \Bigg( &\left\vert \frac{1}{2} , -\frac{1}{2} , \frac{1}{2} , -\frac{1}{2} \right\rangle + \left\vert \frac{1}{2} , -\frac{1}{2} , -\frac{1}{2} , \frac{1}{2} \right\rangle \\
		 - & \left\vert -\frac{1}{2} , \frac{1}{2} , \frac{1}{2} , - \frac{1}{2} \right\rangle - \left\vert -\frac{1}{2} , \frac{1}{2} , -\frac{1}{2} , \frac{1}{2} \right\rangle \Bigg)
		 \end{aligned}
		 \\ \label{eq:c15}
		&\left\vert 0 \, 0 , 1 \, -1 \right\rangle = \frac{1}{\sqrt{2}} \Bigg( \left\vert \frac{1}{2} , -\frac{1}{2} , -\frac{1}{2} , -\frac{1}{2} \right\rangle - \left\vert -\frac{1}{2} , \frac{1}{2} , -\frac{1}{2} , -\frac{1}{2} \right\rangle \Bigg) \\ \label{eq:c16}
		&\left\vert 0 \, 0 , 0 \, 0 \right\rangle = 
		\begin{aligned}[t]
		\frac{1}{2} \Bigg( &\left\vert \frac{1}{2} , -\frac{1}{2} , \frac{1}{2} , -\frac{1}{2} \right\rangle - \left\vert \frac{1}{2} , -\frac{1}{2} , -\frac{1}{2} , \frac{1}{2} \right\rangle \\
		- & \left\vert -\frac{1}{2} , \frac{1}{2} , \frac{1}{2} , - \frac{1}{2} \right\rangle + \left\vert -\frac{1}{2} , \frac{1}{2} , -\frac{1}{2} , \frac{1}{2} \right\rangle \Bigg)
		\end{aligned}
	\end{align}
\end{subequations}
The Clebsch-Gordan coefficients form the orthogonal matrix $\mathcal{Q}$ that implements the basis transformation described above,
\begin{equation}\label{eq:cu}
	\mathcal{Q} =
	\left(
	\begin{array}{cccccccccccccccc}
		1 & 0 & 0 & 0 & 0 & 0 & 0 & 0 & 0 & 0 & 0 & 0 & 0 & 0 & 0 & 0 \\
		0 & \frac{1}{\sqrt{2}} & \frac{1}{\sqrt{2}} & 0 & 0 & 0 & 0 & 0 & 0 & 0 & 0 & 0 & 0 & 0 & 0 & 0 \\
		0 & 0 & 0 & 1 & 0 & 0 & 0 & 0 & 0 & 0 & 0 & 0 & 0 & 0 & 0 & 0 \\
		0 & \frac{1}{\sqrt{2}} & -\frac{1}{\sqrt{2}} & 0 & 0 & 0 & 0 & 0 & 0 & 0 & 0 & 0 & 0 & 0 & 0 & 0 \\
		0 & 0 & 0 & 0 & \frac{1}{\sqrt{2}} & 0 & 0 & 0 & \frac{1}{\sqrt{2}} & 0 & 0 & 0 & 0 & 0 & 0 & 0 \\
		0 & 0 & 0 & 0 & 0 & \frac{1}{2} & \frac{1}{2} & 0 & 0 & \frac{1}{2} & \frac{1}{2} & 0 & 0 & 0 & 0 & 0 \\
		0 & 0 & 0 & 0 & 0 & 0 & 0 & \frac{1}{\sqrt{2}} & 0 & 0 & 0 & \frac{1}{\sqrt{2}} & 0 & 0 & 0 & 0 \\
		0 & 0 & 0 & 0 & 0 & \frac{1}{2} & -\frac{1}{2} & 0 & 0 & \frac{1}{2} & -\frac{1}{2} & 0 & 0 & 0 & 0 & 0 \\
		0 & 0 & 0 & 0 & 0 & 0 & 0 & 0 & 0 & 0 & 0 & 0 & 1 & 0 & 0 & 0 \\
		0 & 0 & 0 & 0 & 0 & 0 & 0 & 0 & 0 & 0 & 0 & 0 & 0 & \frac{1}{\sqrt{2}} & \frac{1}{\sqrt{2}} & 0 \\
		0 & 0 & 0 & 0 & 0 & 0 & 0 & 0 & 0 & 0 & 0 & 0 & 0 & 0 & 0 & 1 \\
		0 & 0 & 0 & 0 & 0 & 0 & 0 & 0 & 0 & 0 & 0 & 0 & 0 & \frac{1}{\sqrt{2}} & -\frac{1}{\sqrt{2}} & 0 \\
		0 & 0 & 0 & 0 & \frac{1}{\sqrt{2}} & 0 & 0 & 0 & -\frac{1}{\sqrt{2}} & 0 & 0 & 0 & 0 & 0 & 0 & 0 \\
		0 & 0 & 0 & 0 & 0 & \frac{1}{2} & \frac{1}{2} & 0 & 0 & -\frac{1}{2} & -\frac{1}{2} & 0 & 0 & 0 & 0 & 0 \\
		0 & 0 & 0 & 0 & 0 & 0 & 0 & \frac{1}{\sqrt{2}} & 0 & 0 & 0 & -\frac{1}{\sqrt{2}} & 0 & 0 & 0 & 0 \\
		0 & 0 & 0 & 0 & 0 & \frac{1}{2} & -\frac{1}{2} & 0 & 0 & -\frac{1}{2} & \frac{1}{2} & 0 & 0 & 0 & 0 & 0
	\end{array}
	\right) ,
\end{equation}
with $\mathcal{Q}^{-1} = \mathcal{Q}^{T}$ corresponding to the inverse transformation.

It is convenient to employ the coupled basis to define the following projection matrices for evaluating the populations of specific molecular-species combinations within the minimal model: $P_{oo, c}$ for both molecules $1$ and $2$ being ortho species, $P_{pp, c}$ for both molecules being para species, and $P_{op, c}$ for the remaining cases, namely molecule $1$ being para and molecule $2$ being ortho, and vice versa. These matrices are diagonal,
\begin{subequations}
	\begin{align} \label{eq:proj1} 
		P_{oo, c} & = \text{diag}(1,1,1,0,1,1,1,0,1,1,1,0,0,0,0,0) \\ \label{eq:proj2} 
		P_{pp, c} & = \text{diag}(0,0,0,0,0,0,0,0,0,0,0,0,0,0,0,1) \\ \label{eq:proj3} 
		P_{op, c} & = \text{diag}(0,0,0,1,0,0,0,1,0,0,0,1,1,1,1,0)  . 
	\end{align}
\end{subequations}
Their sum, $P_{oo, c} + P_{pp, c} + P_{op, c}$, yields the $16 \times 16$ identity matrix. The corresponding matrices in the uncoupled representation are given by $P_{u} = \mathcal{Q} ^{-1} P_{c} \mathcal{Q}$, where $P_{c}$ denotes any of the matrices defined above (see Eqs.~(\ref{eq:proj1})--(\ref{eq:proj3})) and $P_{u}$ is its transformed counterpart.

On the other hand, the matrix representation of the spin components for each of the four protons in the minimal model can be readily obtained in the uncoupled basis. Specifically, the Cartesian spin components are defined as
\begin{subequations}
	\begin{align}
		S_{1 j, u} & = \frac{\hbar}{2} \sigma_{j} \otimes \mathds{1}_{1 B} \otimes \mathds{1}_{2 A} \otimes \mathds{1}_{2 B} \\
		I_{1 j, u} & = \mathds{1}_{1 A} \otimes \frac{\hbar}{2} \sigma_{j} \otimes \mathds{1}_{2 A} \otimes \mathds{1}_{2 B} \\
		S_{2 j, u} & = \mathds{1}_{1 A} \otimes \mathds{1}_{1 B} \otimes \frac{\hbar}{2} \sigma_{j} \otimes \mathds{1}_{2 B} \\
		I_{2 j, u} & = \mathds{1}_{1 A} \otimes \mathds{1}_{1 B} \otimes \mathds{1}_{2 A} \otimes  \frac{\hbar}{2} \sigma_{j} ,
	\end{align}
\end{subequations}
where $j=x,y,z$, $\sigma_{j}$ are the Pauli matrices, and $\mathds{1}_{1 A}$, $\mathds{1}_{1 B}$, $\mathds{1}_{2 A}$, and $\mathds{1}_{2 B}$ denote $2 \times 2$ identity matrices. Explicitly,
\begin{equation}\label{eq:paulisigma1}
	\begin{aligned}
	&\sigma_{x} =
	\begin{pmatrix}
		0 & 1  \\
		1 & 0  
	\end{pmatrix}
	, \quad
	\sigma_{y} =
	\begin{pmatrix}
		0 & -i  \\
		i & 0  
	\end{pmatrix}
	, \quad
	\sigma_{z} =
	\begin{pmatrix}
		1 & 0  \\
		0 & -1  
	\end{pmatrix}
	,
	\\ 
	&\text{and} \quad
	\mathds{1}_{1 A}=\mathds{1}_{1 B} = \mathds{1}_{2 A}=\mathds{1}_{2 B} =
	\begin{pmatrix}
		1 & 0  \\
		0 & 1  
	\end{pmatrix}
	.
	\end{aligned}
\end{equation}
The transformation to the coupled representation is given by $S_{c} = \mathcal{Q} S_{u} \mathcal{Q} ^{-1}$ for $S_{u}$ any of $S_{1 j, u}$, $I_{1 j, u}$, $S_{2 j, u}$, $I_{2 j, u}$.

\section{Additional plots}
\label{app2}

\begin{figure}[h!]
	\centering
	\includegraphics[width=1.0\columnwidth]{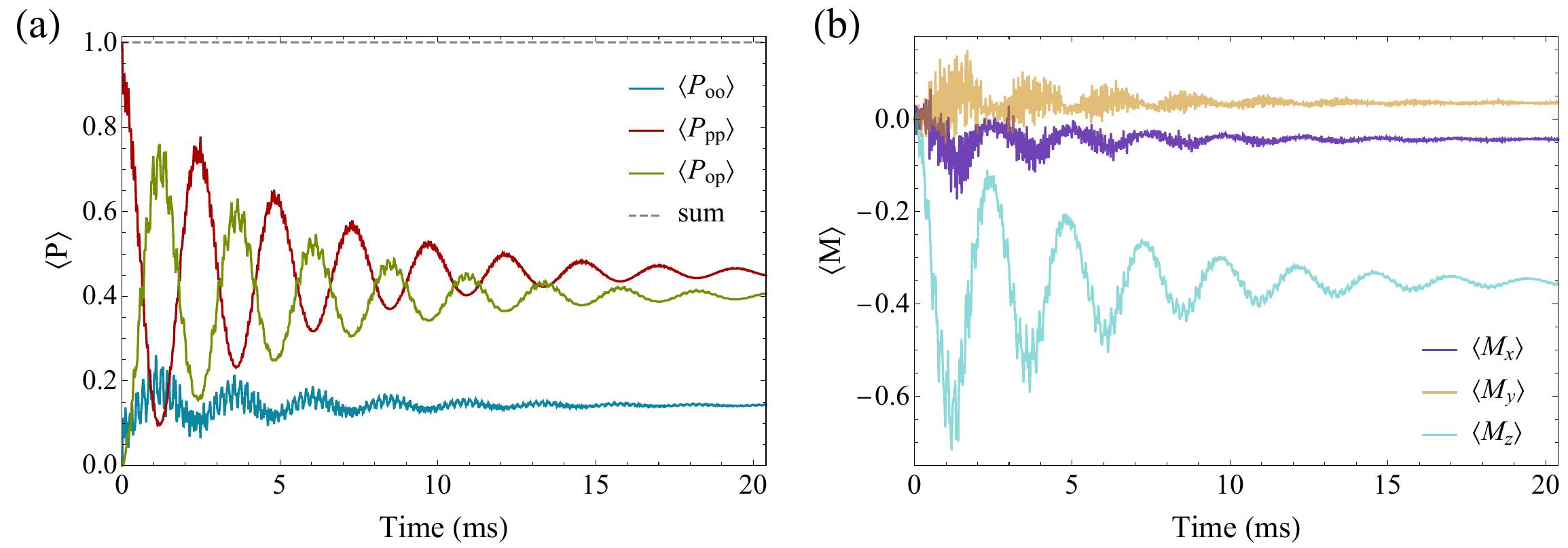}
	\caption{Time evolution of the populations $\left\langle P_{oo}\right\rangle$, $\left\langle P_{pp}\right\rangle$, and $\left\langle P_{op}\right\rangle$ (a), along with the average total spin projection components $\left\langle M_{x} \right\rangle$, $\left\langle M_{y} \right\rangle$, and $\left\langle M_{z} \right\rangle$ (b) for a magnetic field of $0.426$~mT. All molecules are initially prepared in the para state.}\label{figpbt}
\end{figure}

\begin{figure}[h!]\rev{
		\centering
		\includegraphics[width=1.0\columnwidth]{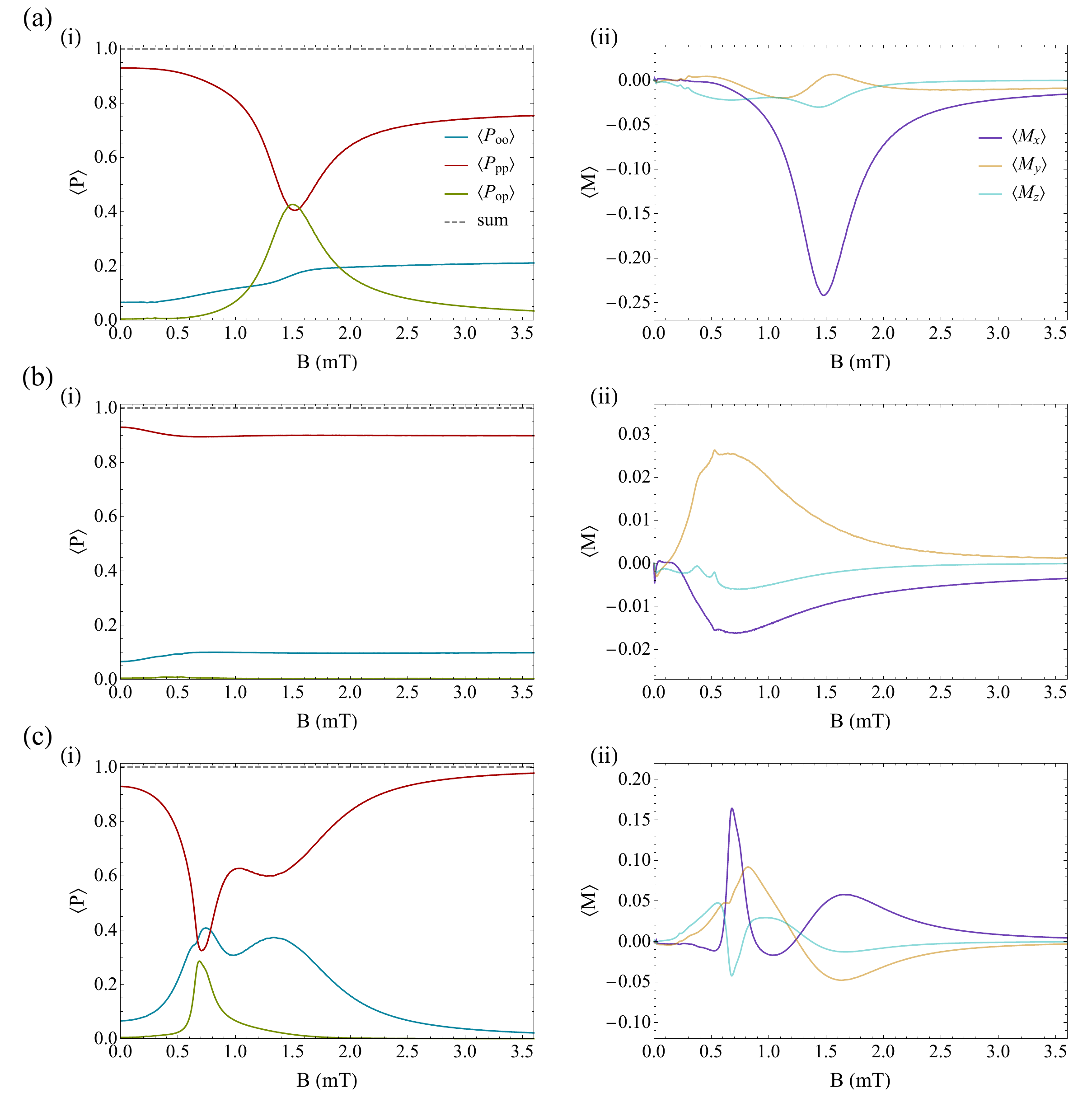}
		\caption{Populations $\left\langle P_{oo}\right\rangle$, $\left\langle P_{pp}\right\rangle$, and $\left\langle P_{op}\right\rangle$ along with the average total spin projection components $\left\langle M_{x} \right\rangle$, $\left\langle M_{y} \right\rangle$, and $\left\langle M_{z} \right\rangle$ at time $6 T_{2}$ as functions of the magnetic field. The molecular geometry is rotated by $45^{\circ}$ (a), $90^{\circ}$ (b), and $135^{\circ}$ (c) about the $y$ axis relative to the direction of the external magnetic field. All molecules are initially prepared in the para state.}\label{figproty}}
\end{figure}

\begin{figure}[h!]\rev{
		\centering
		\includegraphics[width=1.0\columnwidth]{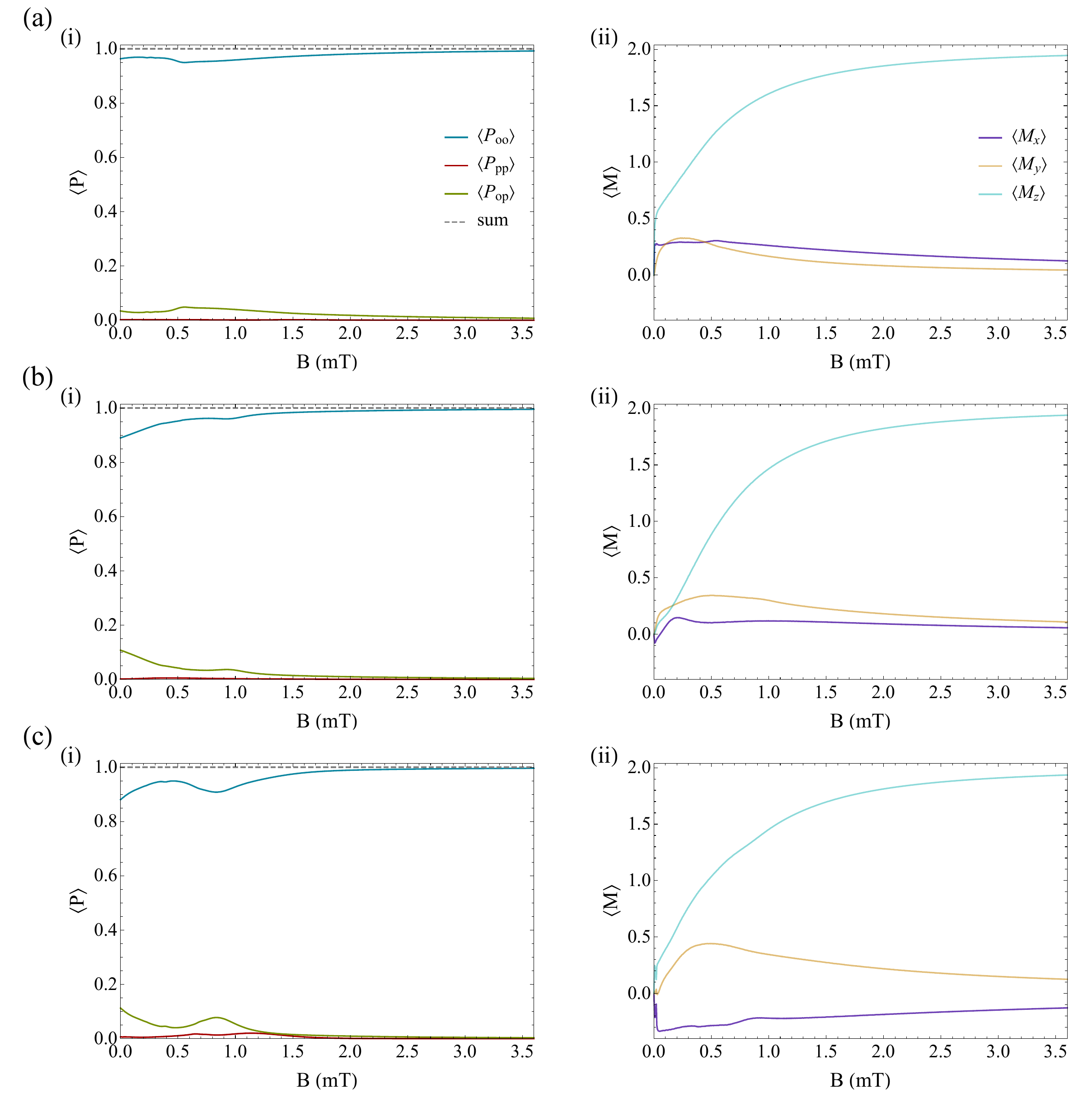}
		\caption{Populations $\left\langle P_{oo}\right\rangle$, $\left\langle P_{pp}\right\rangle$, and $\left\langle P_{op}\right\rangle$ along with the average total spin projection components $\left\langle M_{x} \right\rangle$, $\left\langle M_{y} \right\rangle$, and $\left\langle M_{z} \right\rangle$ at time $6 T_{2}$ as functions of the magnetic field. The molecular geometry is rotated by $45^{\circ}$ (a), $90^{\circ}$ (b), and $135^{\circ}$ (c) about the $x$ axis relative to the direction of the external magnetic field. The initial state corresponds to full spin polarization along the $z$ axis.}\label{figorthroty}}
\end{figure}

\begin{figure}[h!]
	\centering
	\includegraphics[width=1.0\columnwidth]{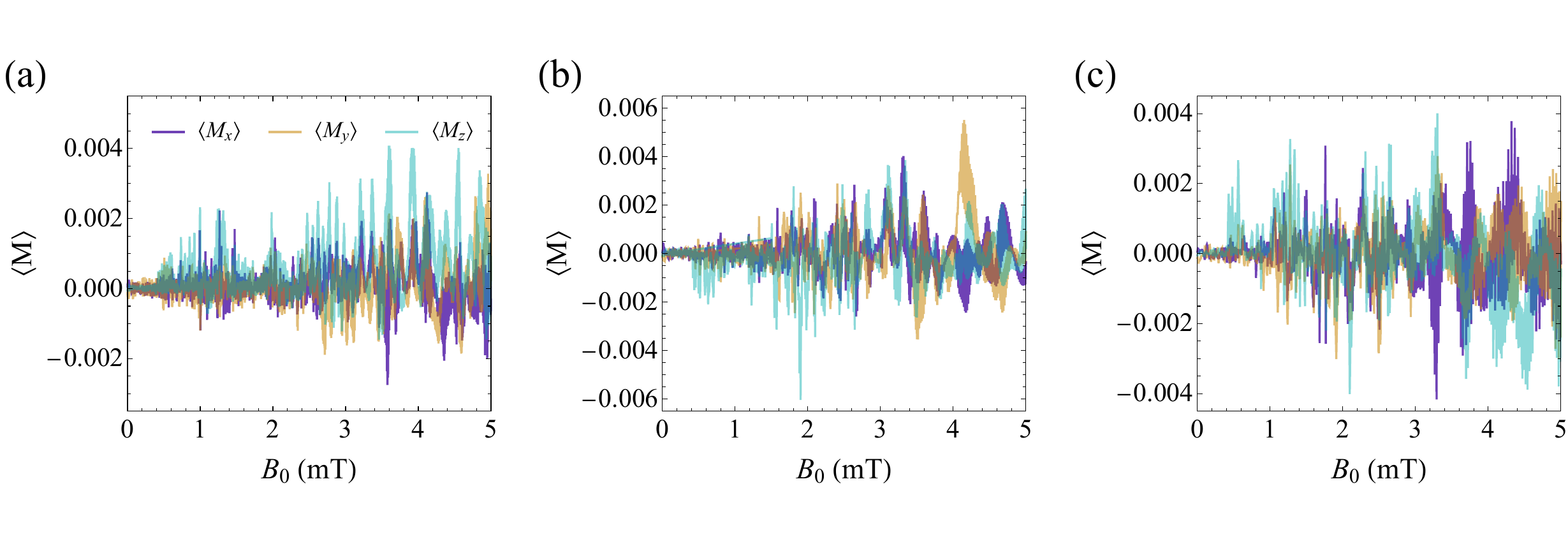}
	\caption{Average total spin projection components $\left\langle M_{x} \right\rangle$, $\left\langle M_{y} \right\rangle$, and $\left\langle M_{z} \right\rangle$ at time $6 T_{2}$ as functions of the magnetic field amplitude $B_{0}$. All molecules are initially prepared in the para state. Each panel corresponds to the magnetic field configuration shown in the respective panel of Fig.~\ref{figpulses}.}\label{figpmabc}
\end{figure}

\clearpage





%


\begin{thebibliography}{10}
	\expandafter\ifx\csname url\endcsname\relax
	\def\url#1{\texttt{#1}}\fi
	\expandafter\ifx\csname urlprefix\endcsname\relax\def\urlprefix{URL }\fi
	\expandafter\ifx\csname href\endcsname\relax
	\def\href#1#2{#2} \def\path#1{#1}\fi
	
	\bibitem{leger1979}
	A.~{Leger}, J.~{Klein}, S.~{de Cheveigne}, C.~{Guinet}, D.~{Defourneau},
	M.~{Belin}, \href{https://ui.adsabs.harvard.edu/abs/1979A&A....79..256L}{{The
			3.1 {\textmu}m absorption in molecular clouds is probably due to amorphous
			H$_{2}$O ice.}}, A \& A 79~(1-2) (1979) 256--259.
	\newline\urlprefix\url{https://ui.adsabs.harvard.edu/abs/1979A&A....79..256L}
	
	\bibitem{weaver1987}
	H.~A. {Weaver}, M.~J. {Mumma}, H.~P. {Larson},
	\href{https://ui.adsabs.harvard.edu/abs/1987A&A...187..411W}{{Infrared
			investigation of water in comet P/Halley}}, A \& A 187~(1-2) (1987) 411--418.
	\newline\urlprefix\url{https://ui.adsabs.harvard.edu/abs/1987A&A...187..411W}
	
	\bibitem{mumma1987}
	M.~J. {Mumma}, H.~A. {Weaver}, H.~P. {Larson},
	\href{https://ui.adsabs.harvard.edu/abs/1987A&A...187..419M}{{The ortho-para
			ratio of water vapor in comet P/ Halley}}, A \& A 187 (1987) 419.
	\newline\urlprefix\url{https://ui.adsabs.harvard.edu/abs/1987A&A...187..419M}
	
	\bibitem{cernicharo1996}
	J.~{Cernicharo}, R.~{Bachiller}, E.~{Gonzalez-Alfonso},
	\href{https://ui.adsabs.harvard.edu/abs/1996A&A...305L...5C}{{Water emission
			at 183 GHz from HH7-11 and other low-mass star-forming regions.}}, A \& A 305
	(1996) L5.
	\newline\urlprefix\url{https://ui.adsabs.harvard.edu/abs/1996A&A...305L...5C}
	
	\bibitem{raich1964}
	J.~C. {Raich}, R.~H. {Good}, Jr., {Ortho-Para Transition in Molecular
		Hydrogen.}, A \& A 139 (1964) 1004.
	\newblock \href {https://doi.org/10.1086/147835} {\path{doi:10.1086/147835}}.
	
	\bibitem{dodelson1986}
	S.~Dodelson, Relativistic treatment of ortho-para-{H}$_2$ transitions, J. Phys.
	B: At. Mol. Phys. 19~(18) (1986) 2871.
	\newblock \href {https://doi.org/10.1088/0022-3700/19/18/017}
	{\path{doi:10.1088/0022-3700/19/18/017}}.
	
	\bibitem{buntkowsky2008}
	G.~Buntkowsky, H.-H. Limbach, B.~Walaszek, A.~Adamczyk, Y.~Xu, H.~Breitzke,
	A.~Schweitzer, T.~Gutmann, M.~Wächtler, N.~Amadeu, D.~Tietze, B.~Chaudret,
	Mechanisms of dipolar ortho/para-{H}$_2${O} conversion in ice, Z. Phys. Chem.
	222~(7) (2008) 1049--1063.
	\newblock \href {https://doi.org/10.1524/zpch.2008.5359}
	{\path{doi:10.1524/zpch.2008.5359}}.
	
	\bibitem{miani2004}
	A.~Miani, J.~Tennyson, Can ortho–para transitions for water be observed?, J.
	Chem. Phys. 120~(6) (2004) 2732--2739.
	\newblock \href {https://doi.org/10.1063/1.1633261}
	{\path{doi:10.1063/1.1633261}}.
	
	\bibitem{meier2018}
	B.~Meier, K.~Kou\ifmmode~\check{r}\else \v{r}\fi{}il, C.~Bengs,
	H.~Kou\ifmmode~\check{r}\else \v{r}\fi{}ilov\'a, T.~C. Barker, S.~J. Elliott,
	S.~Alom, R.~J. Whitby, M.~H. Levitt, Spin-isomer conversion of water at room
	temperature and quantum-rotor-induced nuclear polarization in the
	water-endofullerene {${\mathrm{H}}_{2}\mathrm{O}@{\mathrm{C}}_{60}$}, Phys.
	Rev. Lett. 120 (2018) 266001.
	\newblock \href {https://doi.org/10.1103/PhysRevLett.120.266001}
	{\path{doi:10.1103/PhysRevLett.120.266001}}.
	
	\bibitem{kannis2021}
	C.~S. Kannis, T.~P. Rakitzis, Macroscopic production of highly
	nuclear-spin-polarized molecules from {IR}-excitation and photodissociation
	of molecular beams, Chem. Phys. Lett. 784 (2021) 139092.
	\newblock \href {https://doi.org/10.1016/j.cplett.2021.139092}
	{\path{doi:10.1016/j.cplett.2021.139092}}.
	
	\bibitem{kannis20252}
	C.~S. Kannis, T.~P. Rakitzis, Production of spin-polarized molecular beams via
	microwave or infrared rotational excitation, Chem. Phys. 607 (2026) 113200.
	\newblock \href {https://doi.org/10.1016/j.chemphys.2026.113200}
	{\path{doi:10.1016/j.chemphys.2026.113200}}.
	
	\bibitem{edwards1986}
	C.~M. Edwards, D.~Zhou, N.~S. Sullivan, Unusual low-temperature effects on the
	{NMR} line shapes in solid hydrogen, Phys. Rev. B 34 (1986) 6540--6542.
	\newblock \href {https://doi.org/10.1103/PhysRevB.34.6540}
	{\path{doi:10.1103/PhysRevB.34.6540}}.
	
	\bibitem{ciullo2016}
	G.~Ciullo, R.~Engels, M.~B\"uscher, A.~Vasilyev (Eds.), Nuclear Fusion with
	Polarized Fuel, Springer Proceedings in Physics 187, Springer International
	Publishing, 2016.
	\newblock \href {https://doi.org/10.1007/978-3-319-39471-8}
	{\path{doi:10.1007/978-3-319-39471-8}}.
	
	\bibitem{baylor2023}
	L.~Baylor, A.~Deur, N.~Eidietis, W.~Heidbrink, G.~Jackson, J.~Liu, M.~Lowry,
	G.~Miller, D.~Pace, A.~Sandorfi, S.~Smith, S.~Tafti, K.~Wei, X.~Wei,
	X.~Zheng, Polarized fusion and potential in situ tests of fuel polarization
	survival in a tokamak plasma, Nucl. Fusion 63~(7) (2023) 076009.
	\newblock \href {https://doi.org/10.1088/1741-4326/acc3ae}
	{\path{doi:10.1088/1741-4326/acc3ae}}.
	
	\bibitem{kannis2025}
	C.~S. Kannis, R.~Engels, T.~El-Kordy, N.~Faatz, S.~J. P\"utz, V.~Verhoeven,
	T.~P. Rakitzis, M.~B\"uscher, Spin manipulation and nuclear polarization
	enhancement in particle beams with static magnetic fields, Phys. Rev. A 112
	(2025) 012801.
	\newblock \href {https://doi.org/10.1103/4nr6-xt7m}
	{\path{doi:10.1103/4nr6-xt7m}}.
	
	\bibitem{coudert1992}
	L.~Coudert, Analysis of the rotational levels of water, J. Mol. Spectrosc.
	154~(2) (1992) 427--442.
	\newblock \href {https://doi.org/10.1016/0022-2852(92)90220-I}
	{\path{doi:10.1016/0022-2852(92)90220-I}}.
	
	\bibitem{buntkowsky2025}
	G.~Buntkowsky, {p}ersonal communication (2025).
	
	\bibitem{abragam1961}
	A.~Abragam, The Principles of Nuclear Magnetism, 1st Edition, Oxford at
	Clarendon Press, Oxford, England, 1961.
	
	\bibitem{buntkowsky2006}
	G.~Buntkowsky, B.~Walaszek, A.~Adamczyk, Y.~Xu, H.-H. Limbach, B.~Chaudret,
	Mechanism of nuclear spin initiated para-{H$_2$} to ortho-{H$_2$} conversion,
	Phys. Chem. Chem. Phys. 8 (2006) 1929--1935.
	\newblock \href {https://doi.org/10.1039/B601594H}
	{\path{doi:10.1039/B601594H}}.
	
	\bibitem{buntkowsky1997}
	G.~Buntkowsky, I.~Sack, H.~H. Limbach, B.~Kling, J.~Fuhrhop, Structure
	elucidation of amide bonds with dipolar chemical shift {NMR} spectroscopy, J.
	Phys. Chem. B 101~(51) (1997) 11265--11272.
	\newblock \href {https://doi.org/10.1021/jp971904i}
	{\path{doi:10.1021/jp971904i}}.
	
	\bibitem{tiesinga2021}
	E.~Tiesinga, P.~J. Mohr, D.~B. Newell, B.~N. Taylor, {CODATA} recommended
	values of the fundamental physical constants: 2018, Rev. Mod. Phys. 93 (2021)
	025010.
	\newblock \href {https://doi.org/10.1103/RevModPhys.93.025010}
	{\path{doi:10.1103/RevModPhys.93.025010}}.
	
	\bibitem{modig2002}
	K.~Modig, B.~Halle, Proton magnetic shielding tensor in liquid water, J. Am.
	Chem. Soc. 124~(40) (2002) 12031--12041.
	\newblock \href {https://doi.org/10.1021/ja026981s}
	{\path{doi:10.1021/ja026981s}}.
	
	\bibitem{sofikitis2015}
	D.~Sofikitis, T.~P. Rakitzis, Mesoscopic production of hyperpolarized
	{$^{15}\mathrm{N}_{2}\mathrm{O}$} and {${\mathrm{H}}_{2}\mathrm{O}$} via
	optical excitation, Phys. Rev. A 92 (2015) 032507.
	\newblock \href {https://doi.org/10.1103/PhysRevA.92.032507}
	{\path{doi:10.1103/PhysRevA.92.032507}}.
	
	\bibitem{wise2001}
	T.~Wise, W.~Haeberli, B.~Lorentz, P.~A. Quin, F.~Rathmann, B.~Schwartz, T.~G.
	Walker, A.~Wellinghausen, J.~T. Balewski, J.~Doskow, H.~O. Meyer, R.~E.
	Pollock, B.~v. Przewoski, T.~Rinckel, S.~K. Saha, P.~V. Pancella, Nuclear
	polarization of hydrogen molecules from recombination of polarized atoms,
	Phys. Rev. Lett. 87 (2001) 042701.
	\newblock \href {https://doi.org/10.1103/PhysRevLett.87.042701}
	{\path{doi:10.1103/PhysRevLett.87.042701}}.
	
	\bibitem{engels2015}
	R.~Engels, M.~Gai\ss{}er, R.~Gorski, K.~Grigoryev, M.~Mikirtychyants, A.~Nass,
	F.~Rathmann, H.~Seyfarth, H.~Str\"oher, P.~Weiss, L.~Kochenda, P.~Kravtsov,
	V.~Trofimov, N.~Tschernov, A.~Vasilyev, M.~Vznuzdaev, H.~P.~g. Schieck,
	Production of hyperpolarized {${\mathrm{H}}_{2}$} molecules from
	{$\stackrel{\ensuremath{\rightarrow}}{\mathrm{H}}$} atoms in gas-storage
	cells, Phys. Rev. Lett. 115 (2015) 113007.
	\newblock \href {https://doi.org/10.1103/PhysRevLett.115.113007}
	{\path{doi:10.1103/PhysRevLett.115.113007}}.
	
	\bibitem{kannis2018}
	C.~S. Kannis, G.~E. Katsoprinakis, D.~Sofikitis, T.~P. Rakitzis,
	Nuclear-spin-polarization dynamics of {${\mathrm{H}}_{2}$},
	{${\mathrm{D}}_{2}$}, and {HD} molecules in magnetic fields, Phys. Rev. A 98
	(2018) 043426.
	\newblock \href {https://doi.org/10.1103/PhysRevA.98.043426}
	{\path{doi:10.1103/PhysRevA.98.043426}}.
	
	\bibitem{reistad2004}
	D.~Reistad, B.~Gålnander, T.~Lofnes, Y.-N. Rao, Experiences of operating
	{CELSIUS} with a hydrogen pellet target, Nucl. Instrum. Methods Phys. Res.
	Sect. A 532~(1) (2004) 118--122.
	\newblock \href {https://doi.org/10.1016/j.nima.2004.06.037}
	{\path{doi:10.1016/j.nima.2004.06.037}}.
	
	\bibitem{buntkowsky20062}
	G.~Buntkowsky, H.-H. Limbach, H-solid state {NMR} studies of tunneling
	phenomena and isotope effects in transition metal dihydrides, J. Low Temp.
	Phys. 143~(3) (2006) 55--114.
	\newblock \href {https://doi.org/10.1007/s10909-006-9211-y}
	{\path{doi:10.1007/s10909-006-9211-y}}.
	
\end{thebibliography}

\end{document}